\documentclass[preprint]{ptephy_om}

\preprintnumber{KYUSHU-HET-376} 

\usepackage{graphics}

\usepackage{amsmath} 
\usepackage{amsthm} 
\usepackage{url} 
\usepackage{stmaryrd}
\usepackage{tikz}
\usepackage{tikz-feynhand}
\usepackage{mathtools}
\usepackage{bm}
\usepackage{booktabs}
\usepackage{subcaption}
\usepackage{color}

\DeclareMathOperator{\tr}{tr}

\newcommand{\Slash}[1]{{\ooalign{\hfil/\hfil\crcr$#1$}}}
\numberwithin{equation}{section}

\allowdisplaybreaks

\begin{document}

\title{Gauge-invariant nonperturbative Wilson action in quantum electrodynamics
II}


\author[1]{Sorato Nagao}
\affil[1]{Department of Physics, Kyushu University, 744 Motooka, Nishi-ku,
Fukuoka 819-0395, Japan}

\author[1,2]{Hiroshi Suzuki}
\affil[2]{Quantum and Spacetime Research Institute (QuaSR), Kyushu University,
744 Motooka, Nishi-ku, Fukuoka 819-0395, Japan}





\begin{abstract}%
The gradient flow exact renormalization group (GFERG) is a formulation of the
exact renormalization group for gauge theories that provides a renormalization
group flow preserving the gauge Ward--Takahashi identity exactly. In this
paper, we nonperturbatively solve the GFERG flow equation for the simplest
nontrivial gauge-invariant Wilson action, introduced in a previous paper, for
$D$-dimensional quantum electrodynamics with~$N_f$ flavors (for all~$D$,
flavors are counted in units of four-component spinors). For $D=3$ and~$D=2$,
a nontrivial infrared fixed point exists even for~$N_f=1$, and the structure of
the critical exponents at the fixed point is qualitatively the same as that in
the large-$N_f$ limit. We also obtain the gauge-field propagator in the
continuum limit; for~$D=2$, it reproduces the dynamical mass of the Schwinger
model within a few percent. To the best of our knowledge, apart from the
large-$N_f$ approximation, this is the first example in which the Wilson
renormalization group in a gauge theory is solved nonperturbatively while
manifestly preserving gauge symmetry.
\end{abstract}

\subjectindex{B05,B31,B32}

\maketitle

\section{Introduction}
\label{sec:1}
The Wilson exact renormalization group (ERG), which is based on coarse graining
and rescaling, is a general framework for capturing the behavior of a system
under scale transformations, and its fundamental importance is widely
recognized~\cite{Wilson:1973jj,Morris:1993qb,Becchi:1996an,Pawlowski:2005xe,Igarashi:2009tj,Rosten:2010vm,Dupuis:2020fhh}. Fixed points defined by the ERG and
the critical exponents at them are of great importance, not only because those
notions describe critical phenomena, but also because they are necessary to
define quantum field theory in continuum spacetime. Although the ERG provides a
general viewpoint on these issues, the conventional ERG is based on a momentum
cutoff and cannot manifestly preserve gauge symmetry, which is fundamental in
particle physics.\footnote{%
The gauge or BRST symmetry appears only in a modified
form~\cite{Becchi:1996an}.} In perturbation theory, the requirement of gauge
symmetry can be imposed order by order (see, e.g.,
Ref.~\cite{Igarashi:2009tj}). In nonperturbative applications of the ERG to
gauge theories, including gravity~\cite{Reuter:2019byg}, however, the lack of
manifest gauge symmetry can be a serious problem.

The gradient flow exact renormalization group (GFERG) considered
in~Refs.~\cite{Sonoda:2020vut,Miyakawa:2021hcx,Miyakawa:2021wus,Sonoda:2022fmk,Miyakawa:2023yob,Sonoda:2025tyu,Nagao:2025sri} is a variant of the ERG for gauge
theories that aims to preserve gauge symmetry manifestly.\footnote{%
For the works of Morris et~al.\ with the same motivation, see
Refs.~\cite{Morris:1999px,Morris:2000fs,Arnone:2005vd,Arnone:2005fb,Morris:2005tv}.} See~Refs.~\cite{Abe:2022smm,Miyakawa:2022qbz,Haruna:2023spq} for related works.
The idea of GFERG is based on the observation that the block-spin process in
the ERG for scalar field theory can be understood in terms of heat
diffusion~\cite{Sonoda:2019ibh} (see also Refs.~\cite{Abe:2018zdc,Carosso:2018bmz,Carosso:2018rep,Carosso:2019qpb,Matsumoto:2020lha,Tanaka:2022pwt}). This
observation naturally leads to a formulation of the ERG by a gauge-covariant
diffusion equation, the gradient flow
equation~\cite{Narayanan:2006rf,Luscher:2010iy,Luscher:2011bx,Luscher:2013cpa}.
It turns out that the resulting ERG flow equation preserves gauge or BRST
invariance \emph{manifestly}. It has also been shown that correlation functions
computed with the Wilson action in GFERG coincide with those in the gradient
flow formulation~\cite{Luscher:2010iy,Luscher:2011bx,Luscher:2013cpa} up to
contact terms (see Ref.~\cite{Sonoda:2025tyu} for a readable exposition of this
point).

This paper is a sequel to~Ref.~\cite{Nagao:2026vaf}, hereafter referred to as~I,
in which an application of GFERG to quantum electrodynamics (QED) was studied.
For the spacetime dimensions $D=4$, $3$, and $2$ and the numbers of flavors
$N_f=1$, $2$, $4$, and~$8$, we numerically and nonperturbatively solve the
renormalization group (RG) flow equation for arguably the simplest nontrivial
gauge-invariant ansatz of the one-particle irreducible (1PI) Wilson
action~\cite{Nicoll:1977hi,Wetterich:1992yh,Wetterich:1993ne,Morris:1993qb}
proposed in~I; we determine the fixed points and the critical exponents. The
flow is qualitatively the same as in the large-$N_f$ limit: neither the number
of fixed points nor the number of relevant directions changes. The results
themselves are therefore not very surprising. Nevertheless, to the best of our
knowledge, this is the first example in which the Wilson ERG for a gauge theory
with finite~$N_f$ is solved nonperturbatively \emph{while preserving gauge
symmetry manifestly and exactly}.

The (GF)ERG equation is a nonlinear functional equation and, apart from
perturbation theory, the best we can do is presumably to seek an approximate
solution based on some truncation of the Wilson action. The calculation in
this paper is also an approximate one, based on the particular
ansatz~\eqref{eq:(2.6)}, and it leaves some points unsatisfactory; for
instance, as we will see later, it does not reproduce the universal two-loop
coefficient of the beta function. In the conventional ERG, however, it is
extremely difficult to find an ansatz that satisfies the gauge Ward--Takahashi
(WT) identity, which is itself a nonlinear functional equation, and it is
unclear whether one is really dealing with the correct physical degrees of
freedom. In our GFERG, by contrast, it is quite easy to find an ansatz that
satisfies the gauge WT identity exactly (Eq.~\eqref{eq:(2.6)} being one
example), and GFERG thus has a great advantage in that it incorporates the
correct physical degrees of freedom. In this sense, it is analogous to the
approach based on the block-spin transformation in lattice gauge theory.

This paper is organized as follows. In Sect.~\ref{sec:2}, we recapitulate the
main results of I that are necessary and sufficient for understanding the
present paper; for their derivation, we refer the reader to I.
Section~\ref{sec:3} is the main part of this paper: we numerically solve the
flow equation for the nonperturbative ansatz, illustrate the RG flow,
determine the critical exponents at the fixed points, and obtain the
gauge-field propagator in the continuum limit, which we compare with the exact
result of the Schwinger model for~$D=2$ and with the large-$N_f$ result
for~$D=3$. Section~\ref{sec:4} is devoted to the conclusion, where we briefly
comment on the situation when four-Fermi interactions are included in the
ansatz. In~Appendix~\ref{sec:A}, we summarize the explicit forms of the
quantities that appear in the GFERG flow equation for the ansatz; these
extremely complicated expressions ensure the manifest gauge symmetry.
In~Appendix~\ref{sec:B}, we summarize the computation of the vertex functions
arising from the ansatz.

Throughout this paper, for any spacetime dimension~$D$, we count the number of
flavors~$N_f$ in units of four-component spinors. In terms of two-component
spinors in~$D=3$ and~$D=2$, this corresponds to $N=2N_f$ flavors. In
particular, for~$D=3$, since a four-component spinor consists of a pair of
two-component spinors with opposite parity, the Chern--Simons term is not
induced for integer~$N_f$. In this paper, we present numerical computations for
$N_f=1$, $2$, $4$, and~$8$. We set
\begin{equation}
   \epsilon:=\frac{4-D}{2}.
\end{equation}

\section{Formulation}
\label{sec:2}
In the GFERG formulation of~QED, the ERG equation for the 1PI Wilson
action~$\mathit{\Gamma}_\tau$ is given in momentum space by (this is Eq.~(3.8)
of~I)
\begin{align}
   &\frac{\partial}{\partial\tau}
   \mathit{\Gamma}_\tau[\mathcal{A},\Bar{\Psi},\Psi]
\notag\\
   &=\int_p\,
   \biggl(
   \left[\frac{\delta}{\delta A_\mu(p)}
   +\frac{\delta\mathit{\Gamma}_\tau}{\delta\mathcal{A}_\mu(p)}\right]
   \left[
   \left(2p^2-\epsilon-\gamma_A\right)\mathcal{A}_\mu(p)
   +\frac{\delta\mathit{\Gamma}_\tau}{\delta\mathcal{A}_\mu(-p)}\right]
\notag\\
   &\qquad\qquad\qquad{}
   +\left(\frac{D+2}{2}+p\cdot\frac{\partial}{\partial p}\right)
   \left[\mathcal{A}_\mu(p)
   -\frac{\delta\mathit{\Gamma}_\tau}{\delta\mathcal{A}_\mu(-p)}\right]
   \cdot\frac{\delta\mathit{\Gamma}_\tau}{\delta\mathcal{A}_\mu(p)}
\notag\\
   &\qquad\qquad\qquad{}
   +\left[
   \frac{\delta}{\delta\Bar{\psi}(p)}
   +\frac{\delta}{\delta\Bar{\Psi}(p)}\mathit{\Gamma}_\tau
   \right](-2p^2+\gamma_\psi)
   \Bar{\Psi}(p)
\notag\\
   &\qquad\qquad\qquad{}
   +\left[
   \frac{\delta}{\delta\psi(p)}
   +\frac{\delta}{\delta\Psi(p)}\mathit{\Gamma}_\tau
   \right](-2p^2+\gamma_\psi)
   \Psi(p)
\notag\\
   &\qquad\qquad\qquad{}
   -i\left[
   \frac{\delta}{\delta\psi(p)}
   +\frac{\delta}{\delta\Psi(p)}\mathit{\Gamma}_\tau
   \right]
   \frac{\delta}{\delta\Bar{\Psi}(-p)}\mathit{\Gamma}_\tau
\notag\\
   &\qquad\qquad\qquad{}
   +\left(\frac{D+1}{2}+p\cdot\frac{\partial}{\partial p}\right)
   \left[
   \Psi(p)-i\frac{\delta}{\delta\Bar{\Psi}(-p)}\mathit{\Gamma}_\tau
   \right]
   \cdot\frac{\delta}{\delta\Psi(p)}\mathit{\Gamma}_\tau
\notag\\
   &\qquad\qquad\qquad{}
   +\left(\frac{D+1}{2}+p\cdot\frac{\partial}{\partial p}\right)
   \left[
   \Bar{\Psi}(p)+i\frac{\delta}{\delta\Psi(-p)}\mathit{\Gamma}_\tau
   \right]
   \cdot\frac{\delta}{\delta\Bar{\Psi}(p)}
   \mathit{\Gamma}_\tau
   \biggr)
\notag\\
   &\qquad{}
   +\int_{p,k}\,
   \left[
   \frac{\delta}{\delta\Bar{\psi}(p+k)}
   +\frac{\delta}{\delta\Bar{\Psi}(p+k)}\mathit{\Gamma}_\tau
   \right]
   (-4)p_\mu e_\tau\left[
   \mathcal{A}_\mu(k)
   +\frac{\delta}{\delta A_\mu(-k)}\right]
   \Bar{\Psi}(p)
\notag\\
   &\qquad{}
   +\int_{p,k}\,
   \left[
   \frac{\delta}{\delta\psi(p+k)}
   +\frac{\delta}{\delta\Psi(p+k)}\mathit{\Gamma}_\tau
   \right]
   4p_\mu e_\tau\left[
   \mathcal{A}_\mu(k)
   +\frac{\delta}{\delta A_\mu(-k)}\right]
   \Psi(p)
\notag\\
   &\qquad{}
   +\int_{p,k,l}\,
   \left[
   \frac{\delta}{\delta\Bar{\psi}(p+k+l)}
   +\frac{\delta}{\delta\Bar{\Psi}(p+k+l)}\mathit{\Gamma}_\tau
   \right]
   (-2)e_\tau^2\left[
   \mathcal{A}_\mu(k)
   +\frac{\delta}{\delta A_\mu(-k)}\right]
\notag\\
   &\qquad\qquad\qquad\qquad\qquad\qquad{}
   \times\left[
   \mathcal{A}_\mu(l)
   +\frac{\delta}{\delta A_\mu(-l)}\right]
   \Bar{\Psi}(p)
\notag\\
   &\qquad{}
   +\int_{p,k,l}\,
   \left[
   \frac{\delta}{\delta\psi(p+k+l)}
   +\frac{\delta}{\delta\Psi(p+k+l)}\mathit{\Gamma}_\tau
   \right]
   (-2)e_\tau^2\left[
   \mathcal{A}_\mu(k)
   +\frac{\delta}{\delta A_\mu(-k)}\right]
\notag\\
   &\qquad\qquad\qquad\qquad\qquad\qquad{}
   \times\left[
   \mathcal{A}_\mu(l)
   +\frac{\delta}{\delta A_\mu(-l)}\right]
   \Psi(p).
\label{eq:(2.1)}
\end{align}
In this expression, $\tau$ parametrizes the scale logarithmically and
$\gamma_A$ and~$\gamma_\psi$ are the anomalous dimensions of the gauge and
fermion fields, respectively. $\{\mathcal{A}_\mu,\Psi,\Bar{\Psi}\}$ are the
arguments of the 1PI Wilson action, whereas $\{A_\mu,\psi,\Bar{\psi}\}$ are the
arguments of the (non-1PI) Wilson action. These two sets are related by the
Legendre transformation,
\begin{align}
   A_\mu(x)
   &=\mathcal{A}_\mu(x)
   -\frac{\delta\mathit{\Gamma}_\tau}{\delta\mathcal{A}_\mu(x)},&&
\notag\\
   \psi(x)&=\Psi(x)
   -i\frac{\delta}{\delta\Bar{\Psi}(x)}\mathit{\Gamma}_\tau,&
   \Bar{\psi}(x)&=\Bar{\Psi}(x)
   +i\frac{\delta}{\delta\Psi(x)}\mathit{\Gamma}_\tau.&
\label{eq:(2.2)}
\end{align}
The conventional ERG (with an appropriate momentum cutoff function) would be
obtained by setting~$e_\tau\to0$ in~Eq.~\eqref{eq:(2.1)}. Then, the first
functional derivatives of a field with respect to a field, such
as~$\delta\mathcal{A}_\mu(p)/\delta A_\nu(q)$, contained in that expression are
given by the inverse of the matrix of the second functional derivatives of the
1PI action; this gives the so-called Wetterich equation. The GFERG
equation~\eqref{eq:(2.1)} as it stands depends on~$e_\tau$ and contains higher
functional derivatives of a field with respect to a field. These make the
computation far more complicated, but this is the price for preserving gauge
symmetry manifestly.

Now, it can be shown that the $\tau$-evolution by~Eq.~\eqref{eq:(2.1)}
preserves the following structure,
\begin{equation}
   \mathit{\Gamma}_\tau[\mathcal{A},\Bar{\Psi},\Psi]
   =\mathit{\Gamma}_\tau^{\text{inv}}[\mathcal{A},\Bar{\Psi},\Psi]
   -\frac{1}{2\xi_\tau}\int_k\,
   e^{k^2}\mathcal{A}_\mu(-k)e^{k^2}\mathcal{A}_\nu(k)k_\mu k_\nu,
\label{eq:(2.3)}
\end{equation}
where $\mathit{\Gamma}_\tau^{\text{inv}}$ is the gauge-invariant part, i.e., it is
invariant under the local $U(1)$ gauge transformation,
\begin{align}
   \delta\mathcal{A}_\mu(k)&=ik_\mu\chi(k),&&
\notag\\
   \delta\Psi(p)&=ie_\tau\int_k\,\chi(k)\Psi(p-k),&
   \delta\Bar{\Psi}(-p)
   &=-ie_\tau\int_k\,\chi(k)\Bar{\Psi}(-p-k).
\label{eq:(2.4)}
\end{align}
Note that the last, gauge non-invariant term in~Eq.~\eqref{eq:(2.3)} with
the gauge-fixing parameter $\xi_\tau$ takes a fixed form. The
parameter~$\xi_\tau$, together with~$e_\tau$, obeys the RG flow equations
($\gamma_A$ itself can depend on~$\tau$)
\begin{equation}
   \frac{d}{d\tau}e_\tau^2=-2\gamma_Ae_\tau^2,\qquad
   \frac{d}{d\tau}\xi_\tau=2(\epsilon+\gamma_A)\xi_\tau.
\label{eq:(2.5)}
\end{equation}
These imply $\xi_\tau e_\tau^2e^{-2\epsilon\tau}=\text{const.}$
The structure~\eqref{eq:(2.3)} and the RG equations~\eqref{eq:(2.5)} are
consequences of the gauge WT identity that the GFERG equation~\eqref{eq:(2.1)}
preserves. Since the gauge transformation~\eqref{eq:(2.4)} takes the
\emph{conventional\/} form, it is easy to write down possible local terms
of~$\mathit{\Gamma}_\tau^{\text{inv}}$.

In this paper, following~I, we study the RG flow of
\begin{align}
   \mathit{\Gamma}_\tau[\mathcal{A},\Bar{\Psi},\Psi]
   &=-\frac{1}{2}\int_k\,
   e^{k^2}\mathcal{A}_\mu(-k)e^{k^2}\mathcal{A}_\nu(k)
   \left[
   (k^2\delta_{\mu\nu}-k_\mu k_\nu)\mathcal{K}_\tau(k^2)
   +\frac{1}{\xi_\tau}k_\mu k_\nu
   \right]
\notag\\
   &\qquad{}
   -\int_p\,
   \Bar{\Psi}_{-1}(-p)(\Slash{p}+im_\tau)\Psi_{-1}(p)
   +e_\tau\int_{p,k}\,
   \Bar{\Psi}_{-1}(-p-k)e^{k^2}\Slash{\mathcal{A}}(k)\Psi_{-1}(p),
\label{eq:(2.6)}
\end{align}
where $m_\tau$ is the fermion mass parameter. The
function~$\mathcal{K}_\tau(k^2)$ describes a nontrivial evolution of the
transverse part of the gauge field action. We normalize the gauge potential by
imposing
\begin{equation}
   \mathcal{K}_\tau(k^2=0)=1.
\label{eq:(2.7)}
\end{equation}
Locality, i.e., the analyticity of the function~$\mathcal{K}_\tau(k^2)$,
implies in particular
\begin{equation}
   \left.k^2\mathcal{K}_\tau'(k^2)\right|_{k^2=0}=0.
\label{eq:(2.8)}
\end{equation}

Equation~\eqref{eq:(2.6)} is arguably the simplest nontrivial gauge-invariant
nonperturbative ansatz or truncation for the 1PI Wilson action. Here, field
variables with the subscript~$-1$, the ``$-1$ variables'' first introduced
in~Ref.~\cite{Sonoda:2022fmk}, are defined in terms of the original variables
through the solution of the flow equations (written in coordinate space)
\begin{align}
   \partial_t\mathcal{A}_{t\mu}(x)
   &=\partial_\nu
   \left[
   \partial_\nu\mathcal{A}_{t\mu}(x)-\partial_\mu\mathcal{A}_{t\nu}(x)
   \right]
   +\alpha_0\partial_\mu\partial_\nu\mathcal{A}_{t\nu}(x),&
   \mathcal{A}_{0\mu}(x)&=\mathcal{A}_\mu(x),
\notag\\
   \partial_t\Psi_t(x)
   &=\left\{
   \left[\partial_\mu-ie\mathcal{A}_{t\mu}(x)\right]^2
   +ie\alpha_0\left[\partial_\mu\mathcal{A}_{t\mu}(x)\right]
   \right\}\Psi_t(x),&
   \Psi_0(x)&=\Psi(x),
\notag\\
   \partial_t\Bar{\Psi}_t(x)
   &=\left\{
   \left[\partial_\mu+ie\mathcal{A}_{t\mu}(x)\right]^2
   -ie\alpha_0\left[\partial_\mu\mathcal{A}_{t\mu}(x)\right]
   \right\}\Bar{\Psi}_t(x),&
   \Bar{\Psi}_0(x)&=\Bar{\Psi}(x),
\label{eq:(2.9)}
\end{align}
by solving \emph{backward\/} from~$t=0$ to~$t=-1$. Equation~\eqref{eq:(2.9)} is
nothing but the gradient flow equations
in~Refs.~\cite{Luscher:2010iy,Luscher:2011bx,Luscher:2013cpa} specialized to
QED. In these differential equations, the parameter~$e$ is regarded as a
constant and, after solving the equations from~$t=0$ to~$t=-1$, $e$ is replaced
by~$e_\tau$. For~$\alpha_0=1$, which we set throughout this paper,
$\mathcal{A}_{-1\mu}(x)$ is simply given by
\begin{equation}
   \mathcal{A}_{-1\mu}(x)=e^{-\partial^2}\mathcal{A}_\mu(x)
   \Rightarrow
   \mathcal{A}_{-1\mu}(k)=e^{k^2}\mathcal{A}_\mu(k),
\label{eq:(2.10)}
\end{equation}
whereas $\Psi_{-1}(x)$ and~$\Bar{\Psi}_{-1}(x)$ are of infinite order
in~$\mathcal{A}_\mu(x)$ through Eq.~\eqref{eq:(2.9)}. In~Appendix~\ref{sec:B},
we express $\Psi_{-1}(x)$ and~$\Bar{\Psi}_{-1}(x)$ in terms
of~$\{\mathcal{A}_\mu,\Psi,\Bar{\Psi}\}$. It can also be seen that, for the
gauge transformation~\eqref{eq:(2.4)}, defining the gauge transformation
function~$\chi_t(k)$ by
\begin{equation}
   \partial_t\chi_t(k)=-\alpha_0k^2\chi_t(k),\qquad
   \chi_0(k)=\chi(k),
\label{eq:(2.11)}
\end{equation}
the $-1$ variables transform as
\begin{align}
   \delta\mathcal{A}_{-1\mu}(k)&=ik_\mu\chi_{-1}(k),&&
\notag\\
   \delta\Psi_{-1}(p)&=ie_\tau\int_k\,\chi_{-1}(k)\Psi_{-1}(p-k),&
   \delta\Bar{\Psi}_{-1}(-p)
   &=-ie_\tau\int_k\,\chi_{-1}(k)\Bar{\Psi}_{-1}(-p-k).
\label{eq:(2.12)}
\end{align}
Therefore, the ansatz~\eqref{eq:(2.6)} is, apart from the gauge-fixing term,
invariant under the gauge transformation~\eqref{eq:(2.4)}.

We use the $-1$~variables in our ansatz~\eqref{eq:(2.6)} because the
ansatz~\eqref{eq:(2.6)} reduces to the Gaussian fixed point of the GFERG when
$e_\tau\to0$ and~$\mathcal{K}_\tau\to1$, as shown in~Ref.~\cite{Sonoda:2022fmk}.
We regard this as a minimal requirement for the ansatz of the Wilson action. A
simple gauge-invariant ansatz in terms of the original
variables~$\{\mathcal{A}_\mu,\Psi,\Bar{\Psi}\}$ would not necessarily contain
the Gaussian fixed point in a simple form.

In GFERG, chiral symmetry is realized in the form of the Ginsparg--Wilson
relation~\cite{Ginsparg:1981bj}. Since this relation is nonlinear in the Wilson
action, unfortunately, it is not easy to construct an ansatz that possesses
exact chiral symmetry. Our ansatz~\eqref{eq:(2.6)} is not chirally symmetric
either, even for~$m_\tau=0$, and the mass parameter~$m_\tau$ receives an
additive renormalization; we will see this in the RG flow below.

Now, substituting the ansatz~\eqref{eq:(2.6)} into the GFERG
equation~\eqref{eq:(2.1)}, after a lengthy calculation described in~I, we have,
with~$x:=k^2$,
\begin{align}
   \frac{\partial}{\partial\tau}\mathcal{K}_\tau(x)
   &=-2\left[
   (\epsilon+\gamma_A)\mathcal{K}_\tau(x)+x\mathcal{K}_\tau'(x)
   \right]
   +\mathcal{L}(x),
\notag\\
   \gamma_A&=-\epsilon+\frac{1}{2}\mathcal{L}(0),
\label{eq:(2.13)}
\end{align}
where the function~$\mathcal{L}(k^2)$ is given in~Appendix~\ref{sec:A}; the
latter relation follows from~Eqs.~\eqref{eq:(2.7)} and~\eqref{eq:(2.8)}. Also,
writing the coefficient of the fermion bilinear in the GFERG equation as
\begin{equation}
   \mathcal{R}(p)=a_\tau\Slash{p}+ib_\tau+O(p^2),
\label{eq:(2.14)}
\end{equation}
where the function~$\mathcal{R}(p)$ is also given in~Appendix~\ref{sec:A}, we
have
\begin{equation}
   \frac{d}{d\tau}m_\tau=(1-a_\tau)m_\tau+b_\tau,\qquad
   \gamma_\psi=\frac{1}{2}a_\tau.
\label{eq:(2.15)}
\end{equation}
The RG flow of~$e_\tau$ and~$\xi_\tau$ is given by~Eq.~\eqref{eq:(2.5)}. The 
functions~$\mathcal{L}(k^2)$ and~$\mathcal{R}(p)$ also depend on~$e_\tau$,
$\xi_\tau$, $m_\tau$, and~$\mathcal{K}_\tau(k^2)$. Equations~\eqref{eq:(2.5)},
\eqref{eq:(2.13)}, and~\eqref{eq:(2.15)} form a closed set of coupled
equations. In the next section, we solve this set of equations numerically and
obtain the RG flow.

\section{Numerical solution to the GFERG equation: nonperturbative flow}
\label{sec:3}
From Eqs.~\eqref{eq:(A7)} and~\eqref{eq:(A8)}, the function~$\mathcal{L}(k^2)$
depends linearly on~$\gamma_\psi$. We therefore decompose it as
\begin{equation}
   \mathcal{L}(k^2)=\mathcal{L}_a(k^2)+\gamma_\psi\mathcal{L}_b(k^2).
\label{eq:(3.1)}
\end{equation}
From~Eq.~\eqref{eq:(A9)}, on the other hand, the function~$\mathcal{R}(p)$
depends linearly on $\gamma_A$ and~$\gamma_\psi$. We accordingly write
$a_\tau$ and~$b_\tau$ in~Eq.~\eqref{eq:(2.14)} as
\begin{equation}
   a_\tau=f_0^a+\gamma_Af_A^a+\gamma_\psi f_\psi^a,\qquad
   b_\tau=f_0^b+\gamma_Af_A^b+\gamma_\psi f_\psi^b,
\label{eq:(3.2)}
\end{equation}
where $\mathcal{L}_{a,b}$ and~$f$'s depend on $e_\tau$, $\xi_\tau$, $m_\tau$,
and~$\mathcal{K}_\tau$ but not on $\gamma_A$ or~$\gamma_\psi$. Then,
$\gamma_\psi$ and~$\gamma_A$ at each~$\tau$ are algebraically given by
\begin{equation}
   \gamma_\psi
   =\frac{f_0^a+f_A^a[-\epsilon+\mathcal{L}_a(0)/2]}
   {2-f_A^a\mathcal{L}_b(0)/2-f_\psi^a},\qquad
   \gamma_A=-\epsilon
   +\frac{1}{2}\left[\mathcal{L}_a(0)+\gamma_\psi\mathcal{L}_b(0)\right].
\label{eq:(3.3)}
\end{equation}
In the numerical integration, we use this representation of~$\gamma_\psi$
and~$\gamma_A$.

\begin{figure}[htbp]
\centering
\includegraphics[width=0.9\textwidth]{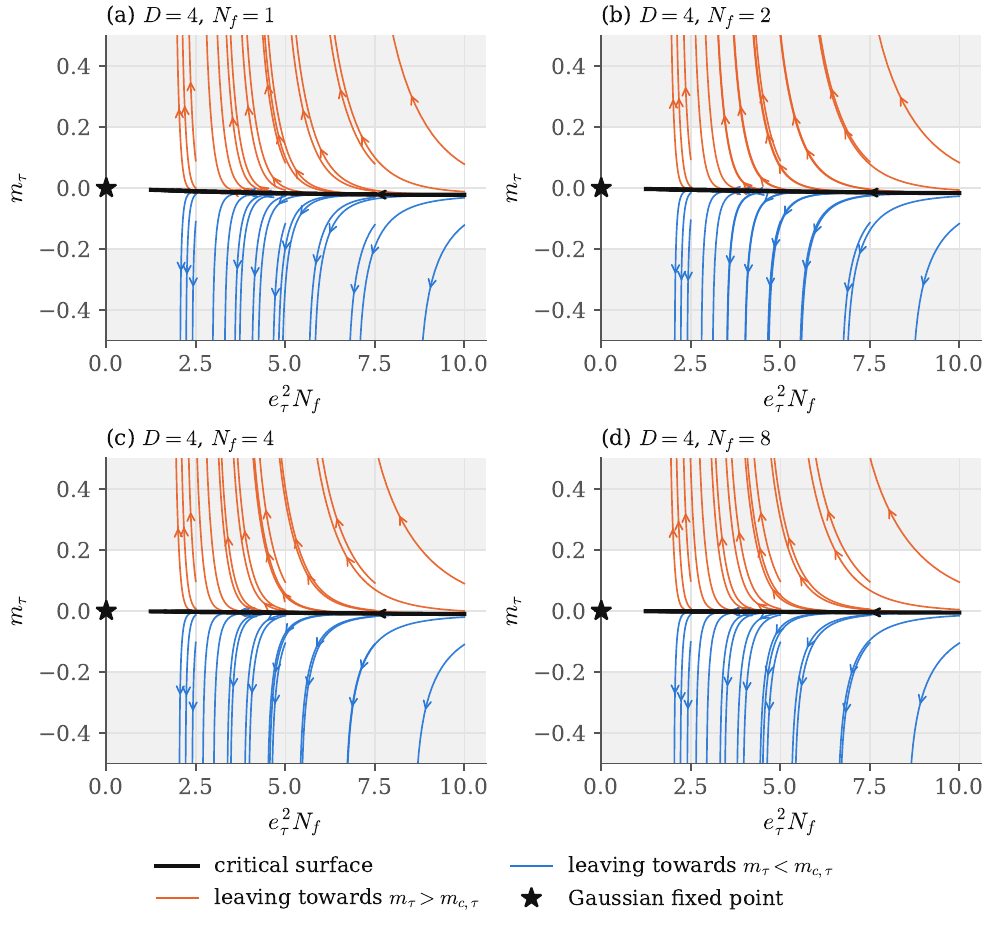}
\caption{Flow toward the infrared (IR), $\tau\to\infty$, projected on the
$(e_\tau^2N_f,m_\tau)$ plane with~$\xi_\tau=0$ for~$D=4$; the number of flavors is
$N_f=1$, $2$, $4$, and~$8$. The thick lines are trajectories on the critical
surface. The star indicates the Gaussian fixed point. The thin lines start from
$m_0=m_{0c}\pm10^{-1}$, \dots, $\pm10^{-8}$; since the flow takes place in an
infinite-dimensional space, projected trajectories may cross. In the shaded
region, $|m_\tau|>0.2$, the interpolation of the two- and three-loop kernels
in~$m_\tau$ becomes an extrapolation.}
\label{fig:1}
\end{figure}

\begin{figure}[htbp]
\centering
\includegraphics[width=0.9\textwidth]{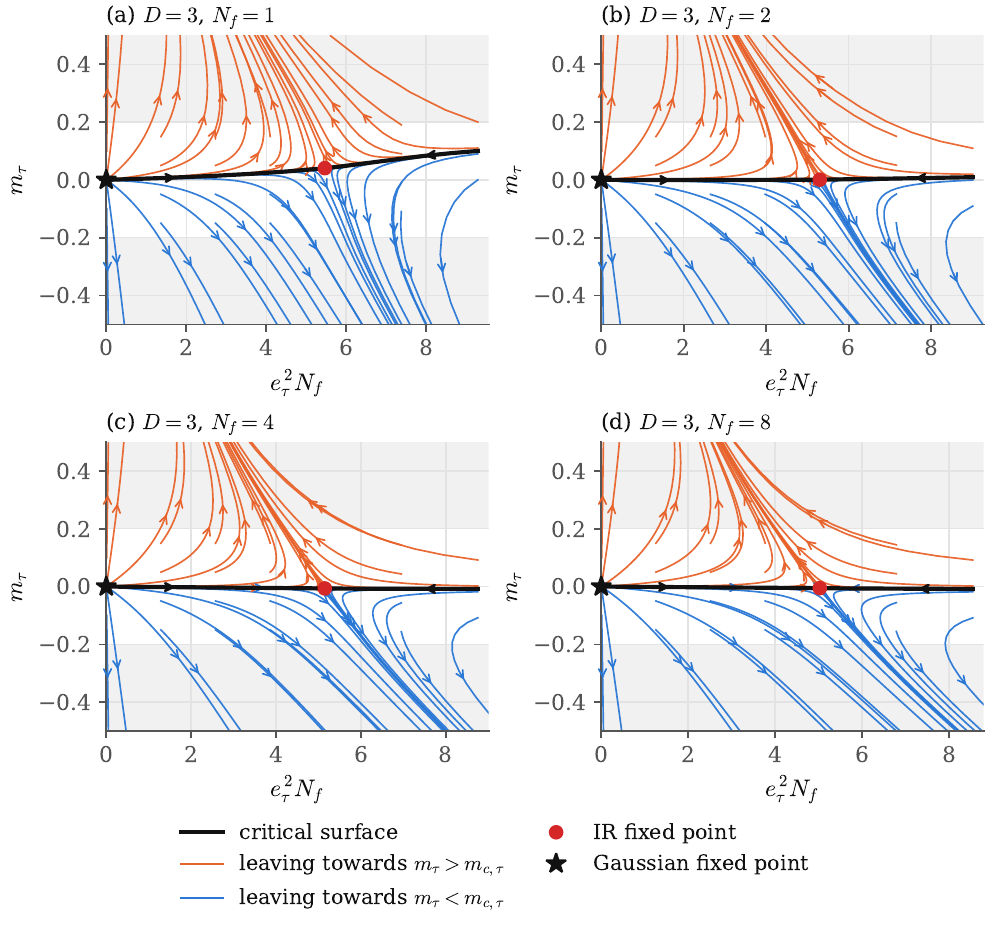}
\caption{The same as Fig.~\ref{fig:1} for~$D=3$. The star indicates the
Gaussian fixed point, whereas the circle indicates the IR fixed point. The
trajectories on the right start from $e_\tau^2=1.7e_*^2$
and~$\mathcal{K}_\tau=\mathcal{K}_*$; some others start from generic points off
the critical surface.}
\label{fig:2}
\end{figure}

\begin{figure}[htbp]
\centering
\includegraphics[width=0.9\textwidth]{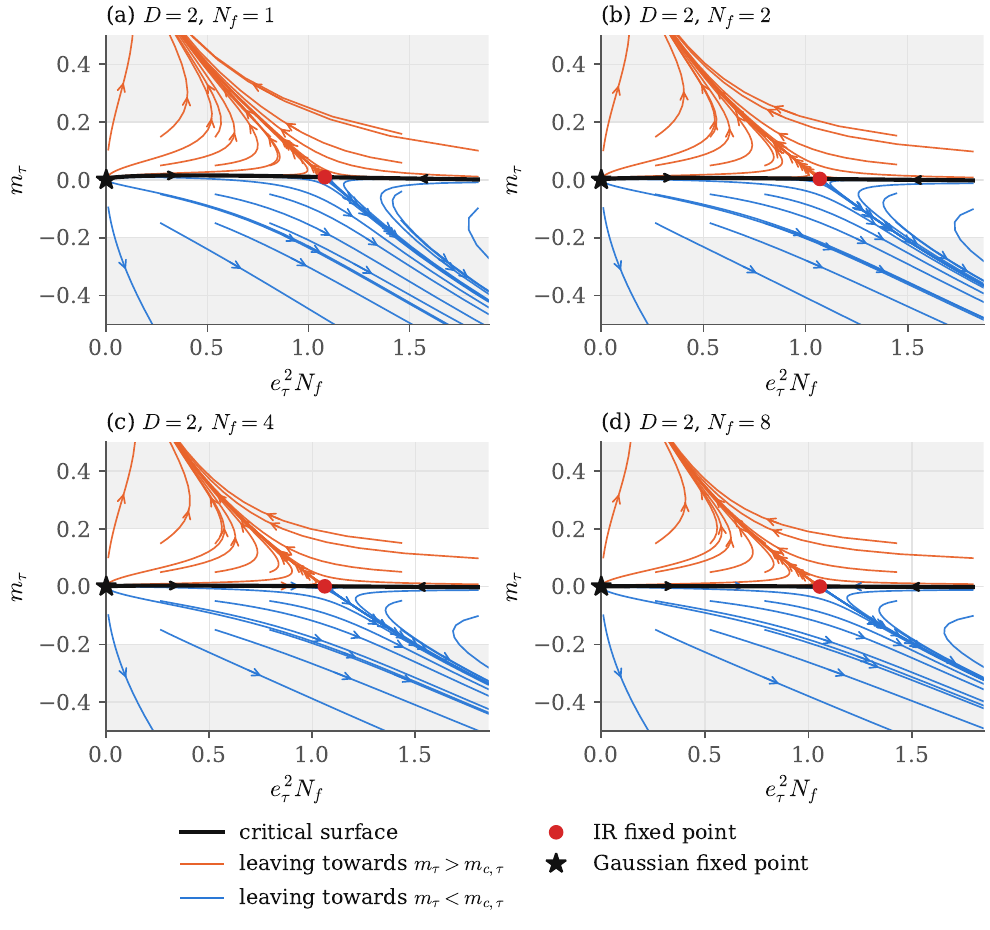}
\caption{The same as Fig.~\ref{fig:1} for~$D=2$. The star indicates the
Gaussian fixed point, whereas the circle indicates the IR fixed point. The
trajectories on the right start from $e_\tau^2=1.7e_*^2$
and~$\mathcal{K}_\tau=\mathcal{K}_*$; some others start from generic points off
the critical surface.}
\label{fig:3}
\end{figure}

The RG flow in the $(e_\tau^2N_f,m_\tau)$ plane with~$\xi_\tau=0$ is plotted
in~Figs.~\ref{fig:1}, \ref{fig:2}, and~\ref{fig:3} for~$D=4$, $3$, and~$2$,
respectively. In each figure, $N_f=1$, $2$, $4$, and~$8$ cases are plotted.

\begin{figure}[htbp]
\centering
\includegraphics[width=0.9\textwidth]{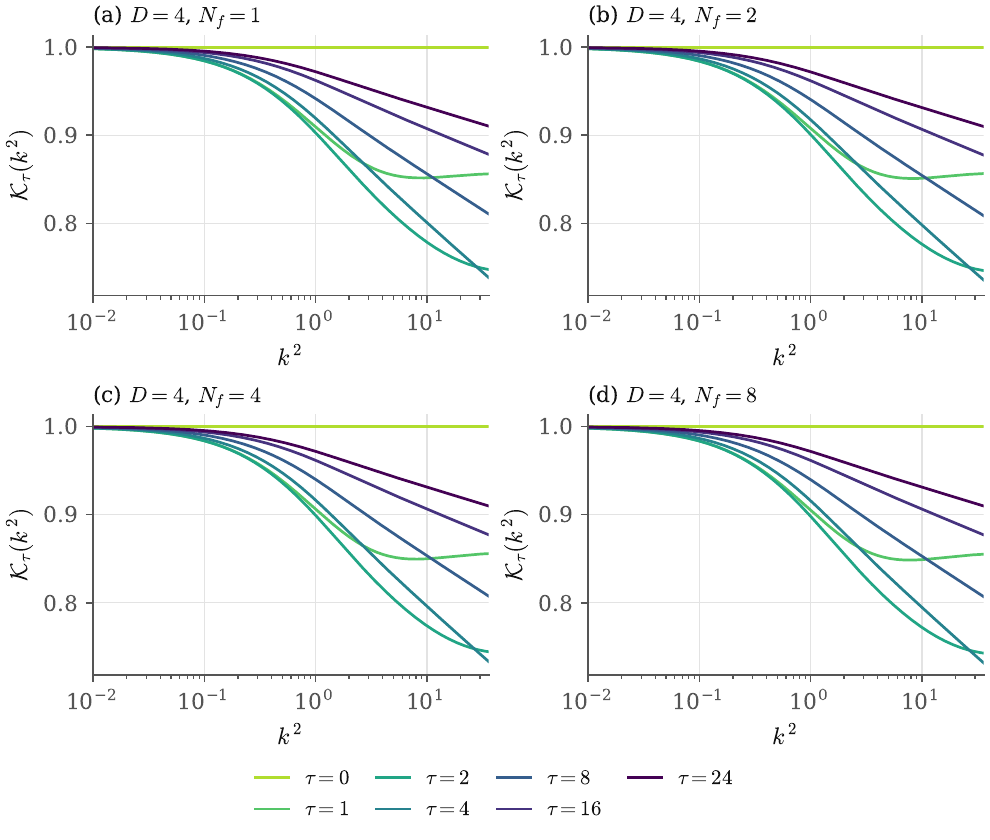}
\caption{The RG evolution of the function~$\mathcal{K}_\tau(k^2)$
in~Eq.~\eqref{eq:(2.6)} along the trajectory on the critical surface for~$D=4$
and $N_f=1$, $2$, $4$, and~$8$.}
\label{fig:4}
\end{figure}

\begin{figure}[htbp]
\centering
\includegraphics[width=0.9\textwidth]{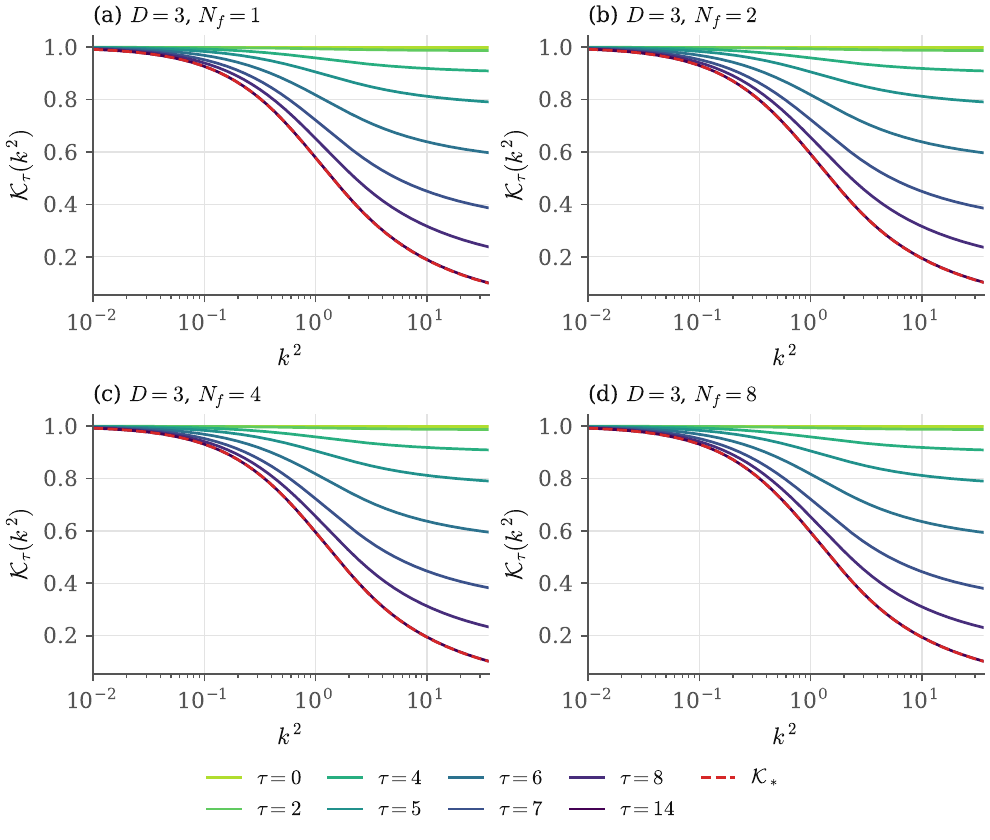}
\caption{The same as~Fig.~\ref{fig:4} but for~$D=3$. The broken line is the
function at the IR fixed point, $\mathcal{K}_*(k^2)$.}
\label{fig:5}
\end{figure}

\begin{figure}[htbp]
\centering
\includegraphics[width=0.9\textwidth]{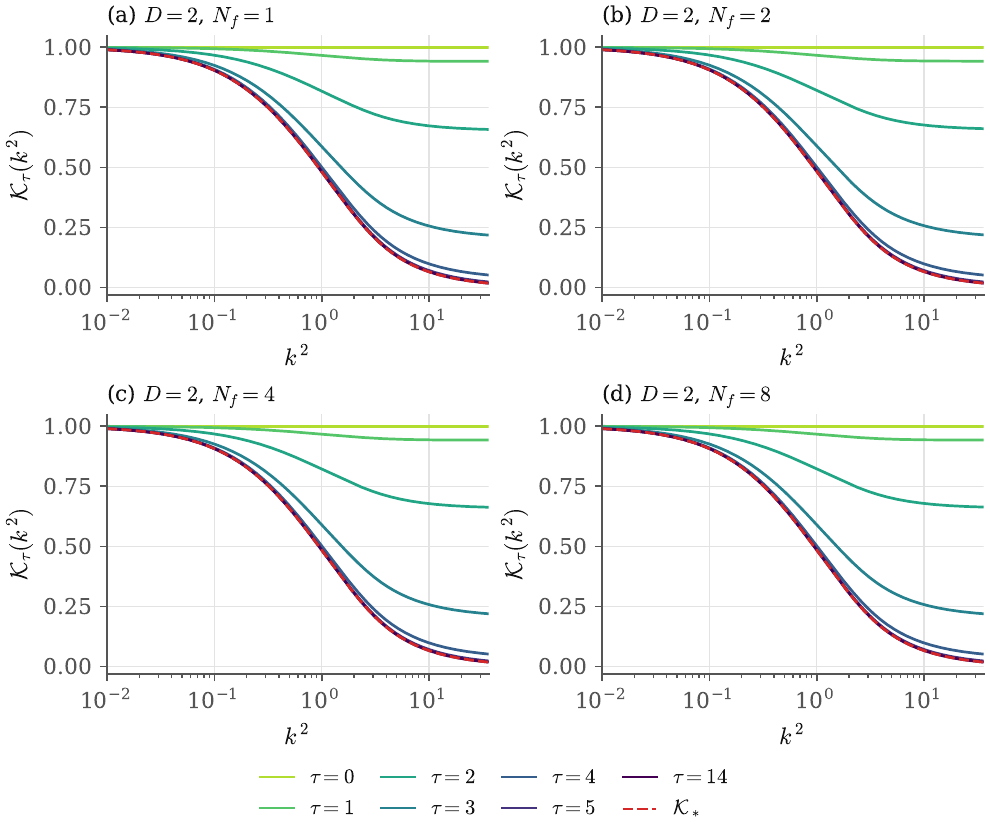}
\caption{The same as~Fig.~\ref{fig:4} but for~$D=2$. The broken line is the
function at the IR fixed point, $\mathcal{K}_*(k^2)$.}
\label{fig:6}
\end{figure}

The RG flow of the function~$\mathcal{K}_\tau(k^2)$ in~Eq.~\eqref{eq:(2.6)}
along the trajectory on the critical surface is plotted in~Figs.~\ref{fig:4},
\ref{fig:5}, and~\ref{fig:6} for~$D=4$, $3$, and~$2$, respectively. In each
figure, $N_f=1$, $2$, $4$, and~$8$ cases are plotted. For $D=3$ and~$2$,
$\mathcal{K}_*(k^2)$ denotes the function at the IR fixed point.

For~$D=4$, i.e., $\epsilon=0$ (Figs.~\ref{fig:1} and~\ref{fig:4}), the only
fixed point is the Gaussian fixed point at~$e_\tau=m_\tau=0$; at that point
$\mathcal{K}_\tau(k^2)=1$. Along the critical surface, the anomalous
dimension~$\gamma_A$ is given by
\begin{equation}
   \gamma_A=\frac{N_fe_\tau^2}{12\pi^2}(1+c_2e_\tau^2),
\label{eq:(3.4)}
\end{equation}
where the numerical calculation shows $c_2=-0.0077$ for~$\xi_\tau=0$. The
one-loop anomalous dimension $N_fe_\tau^2/12\pi^2$ is thus correctly reproduced
(see also~I), but the universal two-loop coefficient, $c_2=3/16\pi^2=0.0190$,
is not reproduced with the ansatz~\eqref{eq:(2.6)}; we attribute this to the
simplicity of the ansatz. We also have, for any~$\xi_\tau$,
\begin{equation}
   \gamma_\psi=\frac{3}{16\pi^2}e_\tau^2+O(e_\tau^4),\qquad
   \frac{d}{d\tau}\ln|m_\tau-m_{c,\tau}|
   =1+\frac{3}{8\pi^2}e_\tau^2+O(e_\tau^4),
\label{eq:(3.5)}
\end{equation}
where $m_{c,\tau}$ denotes the mass parameter on the critical surface and
$|m_\tau-m_{c,\tau}|$ is assumed to be small. These also reproduce the one-loop
results in~Ref.~\cite{Miyakawa:2021wus} (see also~I).\footnote{%
Note that $\gamma_\psi$ is the anomalous dimension of the fermion field
defined through the gradient flow (the $-1$ variables) and thus differs from
the gauge-dependent anomalous dimension of the conventional fermion field.}
However, it turns out that the $N_f$-independent part of the two-loop
coefficient of the mass anomalous dimension, which is scheme independent in
QED, is not reproduced: writing the right-hand side of the second relation
in~Eq.~\eqref{eq:(3.5)} as $1+6a+(\gamma_0'+\gamma_1'N_f)a^2+O(a^3)$ with
$a:=e_\tau^2/16\pi^2$, we find numerically $\gamma_0'\simeq-17$ for~$\xi_\tau=0$,
whereas the universal value is~$3$. From~Eqs.~\eqref{eq:(2.5)},
\eqref{eq:(2.15)}, \eqref{eq:(3.4)}, and~\eqref{eq:(3.5)}, $e_\tau^2$ is
marginally irrelevant, $m_\tau$ is relevant as~Fig.~\ref{fig:1} shows, and
$\xi_\tau$ is marginally relevant.

For $D=3$ ($\epsilon=1/2$) and~$2$ ($\epsilon=1$), as Figs.~\ref{fig:2}
and~\ref{fig:3} show, there exists an IR fixed point as indicated by the
large-$N_f$ approximation~\cite{Appelquist:1988sr}. The ansatz~\eqref{eq:(2.6)}
possesses the IR fixed point even for~$N_f=1$. The critical surface slightly
bends away from~$m_\tau=0$ as a result of an additive mass renormalization
because our ansatz does not preserve chiral symmetry. Note that,
from~Eq.~\eqref{eq:(2.5)}, $\gamma_A=0$ at the fixed point with~$e_*\neq0$, so
that $d\xi_\tau/d\tau=2\epsilon\xi_\tau$; i.e., $\xi_\tau$ is a relevant parameter
with the eigenvalue~$2\epsilon$. Therefore, the fixed point must have
$\xi_\tau=0$, and we set $\xi_\tau=0$ in~Figs.~\ref{fig:2} and~\ref{fig:3}. The
anomalous dimension~$\gamma_\psi$ and the critical exponents at the IR fixed
point are tabulated in~Tables~\ref{table:1} ($D=3$) and~\ref{table:2}
($D=2$).\footnote{%
The anomalous dimension~$\gamma_\psi$ and the critical exponents at the
nontrivial fixed points are independent of the gauge-fixing
parameter~$\xi_\tau$ simply because $\xi_*=0$ at any nontrivial fixed point.
Generally speaking, the gauge WT identity alone is not sufficient to ensure the
$\xi_\tau$-independence of these quantities; one has to impose, in addition,
the Landau--Khalatnikov (LKh) relation~\cite{Sonoda:2013dwa}. It turns out
that, unfortunately, our ansatz~\eqref{eq:(2.6)} does not satisfy the LKh
relation~\cite{Nagao:2026xxx}. In this sense, the $\xi_\tau$-independence of the
results in~Tables~\ref{table:1} and~\ref{table:2} is not guaranteed by symmetry
but merely reflects $\xi_*=0$; in fact, the mass anomalous dimension around the
Gaussian fixed point depends on~$\xi_\tau$ at two-loop order.} The critical
exponents are defined by the flow, Eqs.~\eqref{eq:(2.5)}, \eqref{eq:(2.13)},
and~\eqref{eq:(2.15)}, linearized around the fixed point. Setting $\xi_\tau=0$
and~$(\delta\mathcal{K}_\tau(k^2),\delta e_\tau^2,\delta m_\tau)\propto e^{\lambda\tau}$, we obtain the eigenvalues~$\lambda$. We denote by~$y_m$ the unique positive
eigenvalue, whose eigenvector is dominated by~$\delta m_\tau$.\footnote{%
The correlation length exponent is~$\nu=1/y_m$ and the scaling dimension of the
mass operator is $D-y_m$.} We denote by~$-\omega$ the largest (least negative)
irrelevant eigenvalue, which governs the leading correction to scaling.

\begin{table}[!htbp]
\centering\small
\begin{tabular}{cccccc}\toprule
$N_f$&$e_*^2N_f$&$m_*$&$\gamma_\psi$&$y_m$&$\omega$\\\midrule
$1$ & 5.42(9)   & $+0.039(3)$   & 0.462(24) & 2.277(24) & 0.86(8)\\
$2$ & 5.293(14) & $+0.0009(1)$  & 0.2115(7) & 1.667(2)  & 0.889(8)\\
$4$ & 5.147(3)  & $-0.00652(2)$ & 0.0992    & 1.334(1)  & 0.924(1)\\
$8$ & 5.0282(6) & $-0.00553$    & 0.0476    & 1.164     & 0.954\\
\midrule
$N_f\to\infty$ (extrapolated) & 4.863 & $-0.066/N_f$ & $0.361/N_f$ & $1+1.277/N_f$ & $1-0.464/N_f$\\
\midrule
$1/N_f$ expansion & --- & --- & --- & $1+1.081/N_f$ & ---\\
\bottomrule
\end{tabular}
\caption{The coordinates of the nontrivial IR fixed point, $e_*^2N_f$ and~$m_*$,
the anomalous dimension of the fermion field~$\gamma_\psi$, and the critical
exponents at the fixed point in~$D=3$ ($\xi_\tau=0$, all the terms): $y_m$ is
the relevant eigenvalue associated with the mass and $-\omega$ is the largest
(least negative) irrelevant eigenvalue of the flow linearized around the fixed
point. The numbers in parentheses are the differences between the two halves of
the quasi-Monte Carlo samples of the three-loop kernels. The row
``$N_f\to\infty$ (extrapolated)'' is obtained by the linear extrapolation
in~$1/N_f$ from~$N_f=32$ and~$64$. The row ``$1/N_f$ expansion'' refers to the
result of the standard large-$N_f$ approximation for QED,
$y_m=1+32/(3\pi^2N_f)=1+1.081/N_f$.}
\label{table:1}
\end{table}
\begin{table}[!htbp]
\centering\small
\begin{tabular}{cccccc}\toprule
$N_f$&$e_*^2N_f$&$m_*$&$\gamma_\psi$&$y_m$&$\omega$\\\midrule
$1$ & 1.0729(10) & $+0.00917(7)$ & 0.0508(1) & 2.120(3) & 1.88(4)\\
$2$ & 1.0677(2)  & $+0.00326(1)$ & 0.0342    & 1.516    & 1.900(9)\\
$4$ & 1.0605(1)  & $+0.00087$    & 0.01897   & 1.245    & 1.937(2)\\
$8$ & 1.0548     & $+0.00016$    & 0.00988   & 1.119    & 1.964(1)\\
\midrule
$N_f\to\infty$ (extrapolated) & $1.0472$ & $0$ & $0.0818/N_f$ & $1+0.925/N_f$ & $2-0.323/N_f$\\
\midrule
large-$N_f$ (exact) & --- & --- & --- & $1+1/(2N_f)$ & ---\\
\bottomrule
\end{tabular}
\caption{The same as Table~\ref{table:1} but for~$D=2$. The row
``large-$N_f$ (exact)'' refers to the exact result for the $N$-flavor
Schwinger model in the large-$N_f$ limit. Note that $1.0472=\pi/3$ to the
digits shown.}
\label{table:2}
\end{table}

The function~$\mathcal{K}_\tau(k^2)$ in~Figs.~\ref{fig:5} and~\ref{fig:6} is
physically interesting, because it determines the gauge-field propagator in the
continuum limit. We may go back to the dimensionful formulation of GFERG
(see~I for details): the dimensionless momentum~$k$ and gauge
potential~$\mathcal{A}_\mu(k)$ are related to the dimensionful ones,
$\Tilde{k}$ and~$\Tilde{\mathcal{A}}_\mu(\Tilde{k})$, by
\begin{equation}
   k=\frac{\Tilde{k}}{\Lambda},\qquad
   \mathcal{A}_\mu(k)=\Lambda^{(D+2)/2}\Tilde{\mathcal{A}}_\mu(\Tilde{k}),
\label{eq:(3.6)}
\end{equation}
where $\Lambda=\mu e^{-\tau}$ is the ultraviolet (UV) cutoff of the Wilson
action, i.e., the scale above which momentum modes have been integrated
out, and $\mu$ is an arbitrary mass scale; see Eqs.~(2.32) and~(2.33) of~I.

Now, from~Eq.~\eqref{eq:(2.6)}, the transverse part of the two-point
function of the gauge potential is given by (the average is taken with
respect to the Wilson action~$S_\tau$ obtained by the Legendre transformation
from~${\mit\Gamma}_\tau$; see Eq.~(2.38) of~I)
\begin{equation}
   \left\langle e^{k^2}A_\mu(k)e^{p^2}A_\nu(p)\right\rangle_{S_\tau}
   =\left(\delta_{\mu\nu}-\frac{k_\mu k_\nu}{k^2}\right)
   \frac{1}{k^2\mathcal{K}_\tau(k^2)}
   \delta(k+p)+(\text{contact terms}).
\label{eq:(3.7)}
\end{equation}
Hereafter, ``contact terms'' refer to terms of the form
$\delta_{\mu\nu}f(k^2)\delta(k+p)$ with $f$ an entire function of~$k^2$, which
are local in position space. To find the implication of this in the continuum
limit, we first note the equality
\begin{equation}
   \left\langle\Tilde{A}_\mu(\Tilde{k})\Tilde{A}_\nu(\Tilde{p})
   \right\rangle_{S_\tau}
   =\Lambda^{-2\epsilon}Z_A(\Lambda)^2
   \left\langle A_\mu(\Tilde{t},\Tilde{k})
   A_\nu(\Tilde{t},\Tilde{p})\right\rangle_S
   +(\text{contact terms}),
\label{eq:(3.8)}
\end{equation}
where $A_\mu(k)=\Lambda^{(D+2)/2}\Tilde{A}_\mu(\Tilde{k})$, which follows from
the comparison of Eq.~(3.18) of~Ref.~\cite{Sonoda:2025tyu} and~Eq.~(2.1)
of~I. In this expression, $A_\mu(\Tilde{t},\Tilde{k})$ on the right-hand side
is the flowed gauge potential in the gradient flow formalism and $S$ is the
bare action; from~Eq.~(2.2) of~I, $\Tilde{t}=1/\Lambda^2$
for~$\Lambda_0\to\infty$. For~$\alpha_0=1$, the flowed gauge potential is
given by $A_\mu(\Tilde{t},\Tilde{k})=e^{-\Tilde{t}\Tilde{k}^2}A_\mu(\Tilde{k})$ in
terms of the bare gauge potential~$A_\mu(\Tilde{k})$. Noting further that
$e_\tau=Z_A(\Lambda)^{-1}$ [Eq.~(2.35) of~I] and
$\Tilde{t}\Tilde{k}^2=\Tilde{k}^2/\Lambda^2=k^2$, the above relation can be
written as
\begin{equation}
   \Lambda^{2\epsilon}e_\tau^2
   \left\langle
   e^{k^2}\!\Tilde{A}_\mu(\Tilde{k})
   e^{p^2}\!\Tilde{A}_\nu(\Tilde{p})
   \right\rangle_{S_\tau}
   =\left\langle A_\mu(\Tilde{k})A_\nu(\Tilde{p})\right\rangle_S,
\label{eq:(3.9)}
\end{equation}
up to contact terms. The right-hand side of this relation is independent
of~$\Lambda$ and~$\tau$. We thus define the gauge-field propagator by the
left-hand side,
\begin{align}
   &D_{\mu\nu}(\Tilde{k})\delta(\Tilde{k}+\Tilde{p})
\notag\\
   &:=\Lambda^{2\epsilon}e_\tau^2
   \left\langle
   e^{k^2}\!\Tilde{A}_\mu(\Tilde{k})
   e^{p^2}\!\Tilde{A}_\nu(\Tilde{p})
   \right\rangle_{S_\tau}
\notag\\
   &=\left(\delta_{\mu\nu}-\frac{\Tilde{k}_\mu\Tilde{k}_\nu}{\Tilde{k}^2}\right)
   \frac{e_\tau^2}
   {(\Tilde{k}^2/\Lambda^2)^\epsilon\mathcal{K}_\tau(\Tilde{k}^2/\Lambda^2)}
   \frac{1}{(\Tilde{k}^2)^{1-\epsilon}}
   \delta(\Tilde{k}+\Tilde{p}),
\label{eq:(3.10)}
\end{align}
up to contact terms, where we have used~Eq.~\eqref{eq:(3.7)} and
$\delta(k+p)=\Lambda^D\delta(\Tilde{k}+\Tilde{p})$. We write
$D_{\mu\nu}(\Tilde{k})=(\delta_{\mu\nu}-\Tilde{k}_\mu\Tilde{k}_\nu/\Tilde{k}^2)D(\Tilde{k})$ in what follows.

Along the critical trajectory, $e_\tau\to e_*$
and~$\mathcal{K}_\tau\to\mathcal{K}_*$ as $\Lambda\to0$. Note, however, that the
latter is the convergence at a fixed dimensionless momentum~$k$, whereas
in~Eq.~\eqref{eq:(3.10)} the argument $k^2=\Tilde{k}^2/\Lambda^2$ also goes to
infinity. Considering the $\tau$-evolution in the flow
equation~\eqref{eq:(2.13)} to larger~$\tau$ (corresponding to
lowering~$\Lambda$), the term $-2x\mathcal{K}_\tau'(x)$ merely transports the
value of~$\mathcal{K}_\tau$ at a fixed dimensionful momentum~$\Tilde{k}$,
whereas the loop corrections are effective only for $k^2\lesssim O(1)$, i.e.,
$\Tilde{k}\lesssim\Lambda$. Thus, $D(\Tilde{k})$ is essentially determined when
$\Lambda\sim\Tilde{k}$ and is frozen afterward. For $\Tilde{k}$ much smaller
than the dynamical scale set by the dimensionful gauge coupling,
$|\Tilde{k}|\ll e(\mu)^{1/\epsilon}$, this happens after the flow has approached
the IR fixed point, and $\mathcal{K}_\tau$ in~Eq.~\eqref{eq:(3.10)} can be
replaced by its fixed-point form~$\mathcal{K}_*$. Therefore, in this IR region,
Eq.~\eqref{eq:(3.10)} possesses a finite continuum limit if
$(k^2)^\epsilon\mathcal{K}_*(k^2)$ tends to a constant as~$k^2\to\infty$. We
numerically find that this is indeed the case for both $D=3$ and~$2$
(see~Table~\ref{table:3} below). Setting
\begin{equation}
   c:=\lim_{k^2\to\infty}(k^2)^\epsilon\mathcal{K}_*(k^2),
\label{eq:(3.11)}
\end{equation}
we have
\begin{equation}
   D(\Tilde{k})
   \stackrel{\Lambda\to0}{\longrightarrow}
   \frac{e_*^2}{c}\frac{1}{(\Tilde{k}^2)^{1-\epsilon}},\qquad
   |\Tilde{k}|\ll e(\mu)^{1/\epsilon}.
\label{eq:(3.12)}
\end{equation}
The dependence on~$\Lambda$ cancels out, and the propagator has a finite
continuum limit determined by the single number~$e_*^2/c$. The
power~$1-\epsilon$ is the scaling dimension of the gauge potential at the IR
fixed point (see also Eq.~(4.33) of~I).

\begin{figure}[htbp]
\centering
\includegraphics[width=0.9\textwidth]{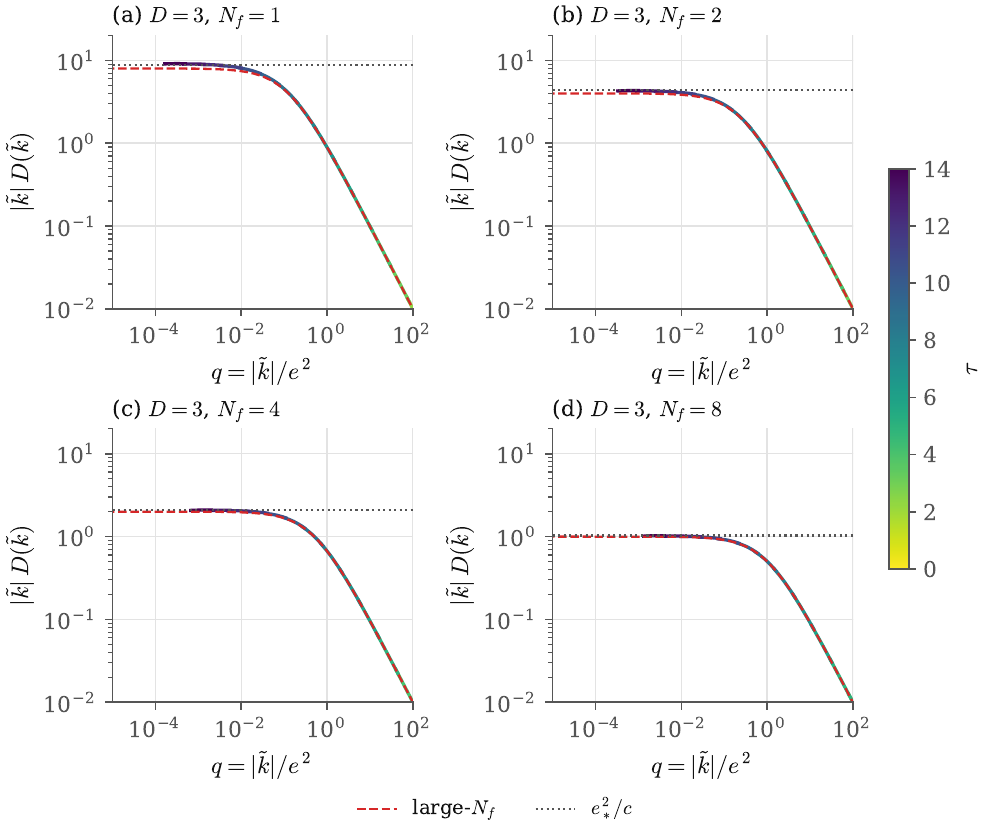}
\caption{The continuum gauge-field propagator for~$D=3$ and $N_f=1$, $2$, $4$,
and~$8$, multiplied by~$|\Tilde{k}|$, as a function of $q=|\Tilde{k}|/e^2$,
where $e=e(\mu)$ is the dimensionful coupling at~$\tau=0$. Each colored segment
is Eq.~\eqref{eq:(3.10)} evaluated at a given~$\tau$ in the window
$4\le\Tilde{k}^2/\Lambda^2\le36$; the color indicates~$\tau$. The broken line
is the large-$N_f$ result $|\Tilde{k}|D(\Tilde{k})=1/(q+N_f/8)$, and the dotted
line is the IR limit~$e_*^2/c$ in~Eq.~\eqref{eq:(3.12)}.}
\label{fig:7}
\end{figure}

\begin{figure}[htbp]
\centering
\includegraphics[width=0.9\textwidth]{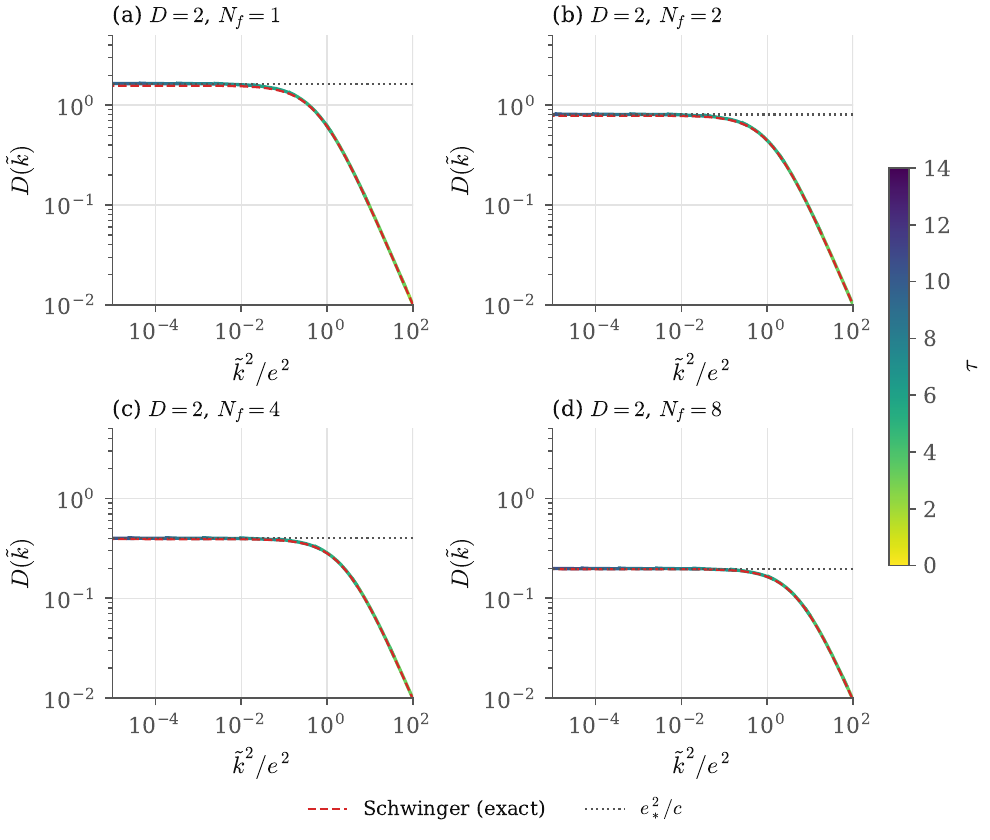}
\caption{The continuum gauge-field propagator for~$D=2$ and $N_f=1$, $2$, $4$,
and~$8$, as a function of~$\Tilde{k}^2/e^2$, where $e=e(\mu)$. The segments are
obtained as in~Fig.~\ref{fig:7}. The broken line is the exact result of the
Schwinger model, $D(\Tilde{k})=1/(\Tilde{k}^2/e^2+2N_f/\pi)$, and the dotted
line is the IR limit~$e_*^2/c$ in~Eq.~\eqref{eq:(3.12)}.}
\label{fig:8}
\end{figure}

The whole momentum dependence of the continuum propagator, including the UV
region $|\Tilde{k}|\gtrsim e(\mu)^{1/\epsilon}$, can be obtained by following the

above picture literally. At each~$\tau$ along the critical trajectory, we
evaluate Eq.~\eqref{eq:(3.10)} in the window $4\le k^2\le36$, i.e., for momenta
just above the cutoff~$\Lambda$, which have been integrated out. As $\tau$
increases, this window moves toward smaller~$\Tilde{k}$, and the segments
obtained at successive values of~$\tau$ join into a single curve. The result is
shown in~Figs.~\ref{fig:7} ($D=3$) and~\ref{fig:8} ($D=2$), where the flow
starts from $e_\tau^2N_f=0.01$ at~$\tau=0$ and $e$ denotes the dimensionful
coupling at~$\tau=0$, $e(\mu)=\mu^\epsilon e_{\tau=0}$. The resulting propagator
interpolates between the free behavior~$e^2/\Tilde{k}^2$ in the UV region and
the IR behavior~\eqref{eq:(3.12)}.

The result can be compared with known results. For~$D=2$,
Eq.~\eqref{eq:(3.12)} does not depend on the momentum, which shows that the
gauge boson acquires a dynamical mass. In $D=2$ QED, i.e., the Schwinger model,
the exact solution gives the propagator $D(\Tilde{k})=e^2/(\Tilde{k}^2+m^2)$
with the dynamical mass squared $m^2=e^2N/\pi=2N_fe^2/\pi$, where $N=2N_f$ is
the number of two-component fermions. For~$\Tilde{k}^2\ll m^2$, the propagator
becomes~$\pi/(2N_f)$, which is independent of~$e$ and is to be compared
with~$e_*^2/c$. For~$D=3$, in the large-$N_f$ approximation, the vacuum
polarization is~$N_fe^2|\Tilde{k}|/8$~\cite{Appelquist:1988sr}, so that
$D(\Tilde{k})=e^2/(\Tilde{k}^2+N_fe^2|\Tilde{k}|/8)$, which behaves as
$8/(N_f|\Tilde{k}|)$ for~$|\Tilde{k}|\ll e^2$; this is to be compared with
$(e_*^2/c)/|\Tilde{k}|$. These reference propagators are also plotted
in~Figs.~\ref{fig:7} and~\ref{fig:8}. For~$D=2$, the propagator obtained from
the ansatz~\eqref{eq:(2.6)} is almost indistinguishable from the exact one over
the whole range of momentum, except for a slight deviation in the IR region
for~$N_f=1$. Table~\ref{table:3} summarizes the comparison in the IR region.
The ansatz~\eqref{eq:(2.6)} reproduces the exact result for~$D=2$ within a few
percent and the large-$N_f$ result for~$D=3$ within about ten percent. In both
cases, the deviation decreases as~$1/N_f$. In the large-$N_f$ limit (see~I),
$c=2/3$ and $e_*^2N_f=\pi/3$ for~$D=2$, and $c=0.608$ and $e_*^2N_f=4.862$
for~$D=3$, so that the agreement becomes exact.
\begin{table}[htbp]
\centering
\begin{tabular}{ccccccc}
\toprule
 & \multicolumn{3}{c}{$D=3$} & \multicolumn{3}{c}{$D=2$}\\
\cmidrule(lr){2-4}\cmidrule(lr){5-7}
$N_f$ & $c$ & $N_fe_*^2/c$ & ratio to~$8$
      & $c$ & $N_fe_*^2/c$ & ratio to~$\pi/2$\\
\midrule
1 & $0.61(3)$   & $8.9(3)$   & $1.11(4)$   & $0.6558(5)$ & $1.636(3)$  & $1.042(2)$\\
2 & $0.6096(1)$ & $8.68(1)$  & $1.085(2)$  & $0.6621(1)$ & $1.6124(6)$ & $1.0265(4)$\\
4 & $0.6120(4)$ & $8.410$    & $1.051$     & $0.6651$    & $1.5946(2)$ & $1.0152(1)$\\
8 & $0.6112(1)$ & $8.226$    & $1.028$     & $0.6661$    & $1.5836$    & $1.0082$\\
\midrule
$N_f\to\infty$ (I) & $0.608$ & $8.00$ & $1$ & $2/3$ & $\pi/2$ & $1$\\
\bottomrule
\end{tabular}
\caption{The constant~$c$ in~Eq.~\eqref{eq:(3.11)} and the coefficient
$N_fe_*^2/c$ of the continuum gauge-field propagator~\eqref{eq:(3.12)} at the
IR fixed point. For~$D=2$, the reference value~$\pi/2$ is the exact result of
the Schwinger model; for~$D=3$, the reference value~$8$ is the result of the
leading large-$N_f$ approximation~\cite{Appelquist:1988sr}. The numbers in
parentheses are estimated as in~Table~\ref{table:1}. The last row is the
leading large-$N_f$ result in~I.}
\label{table:3}
\end{table}

\clearpage

\section{Conclusion}
\label{sec:4}
In this paper, based on the results of our previous paper~\cite{Nagao:2026vaf},
we have further developed the application of GFERG, the Wilson exact
renormalization group that manifestly preserves gauge symmetry, to QED. Assuming
the 1PI Wilson action~\eqref{eq:(2.6)}, which satisfies the gauge WT identity,
we have numerically obtained its nonperturbative RG flow. Along this RG flow,
the gauge WT identity is exactly preserved. For all of $D=4$, $3$, and~$2$, the
results are qualitatively the same as the picture obtained by perturbation
theory or the large-$N_f$ approximation, as far as the existence of fixed
points and the number of relevant directions are concerned. In this sense, the
results themselves are not surprising. Nevertheless, to the best of our
knowledge, this is the first nonperturbative RG flow that exactly preserves
gauge symmetry, which guarantees the correct physical degrees of freedom in
gauge theory. We have also obtained the gauge-field propagator in the continuum
limit; it reproduces the dynamical mass of the Schwinger model for~$D=2$ within
a few percent and the large-$N_f$ infrared behavior for~$D=3$ within about ten
percent, and the deviations decrease as~$1/N_f$.

The gauge theory considered in this paper is the simplest one, and extensions
to more complicated systems are desirable. Within $U(1)$ gauge theory, the
first possibility would be to add four-Fermi interactions, for example,
\begin{align}
   {\mit\Gamma}_\tau^{\text{4Fermi}}
   &:=-\int d^Dx\,
   \frac{G_\tau^S}{2}\left\{
   \left[\Bar{\Psi}_{-1}(x)\Psi_{-1}(x)\right]^2
   -\left[\Bar{\Psi}_{-1}(x)\gamma_5\Psi_{-1}(x)\right]^2
   \right\}
\notag\\
   &\qquad{}
   -\int d^Dx\,
   \frac{G_\tau^V}{2}\left\{
   \left[\Bar{\Psi}_{-1}(x)\gamma_\mu\Psi_{-1}(x)\right]^2
   +\left[\Bar{\Psi}_{-1}(x)\gamma_5\gamma_\mu\Psi_{-1}(x)\right]^2
   \right\},
\label{eq:(4.1)}
\end{align}
to Eq.~\eqref{eq:(2.6)}. The nonperturbative RG flow of this system has been
studied for a long time~\cite{Aoki:1996fh,Gies:2004hy,Igarashi:2016gcf,Gies:2020xuh,Igarashi:2021zml,Echigo:2025dia,Cresswell-Hogg:2026jpe}, and a comparison
with those studies is desirable.\footnote{%
A preliminary study indicates that, for $D=4$ and~$N_f=1$, the 1PI Wilson
action given by Eq.~\eqref{eq:(2.6)} plus~Eq.~\eqref{eq:(4.1)} possesses only
the Gaussian fixed point and the NJL fixed point (at which $e_*=0$), apart from
presumably unphysical fixed points at which $|G_*^S|$ or~$|G_*^V|$ is
of~$O(10^3)$ or larger. We will report this study
elsewhere~\cite{Nagao:2026yyy}.} We also aim at extensions to non-Abelian gauge
theories and further to gravity.

\section*{Acknowledgments}
We would like to thank Hidenori Sonoda for many years of valuable discussions.
The work of H.S. was supported by JSPS KAKENHI Grant Number JP23739313 and by
the MEXT program ``Supporting Pioneering Research through AI for 1,000
Discovery Challenges (SPReAD),'' Grant Number JPMXP1726257479. It was also
partially supported by the Quantum and Spacetime Research Institute, Kyushu
University.
We received substantial assistance from Claude (Anthropic) in carrying out this
study. Needless to say, any remaining errors are our own.

\appendix

\section{Explicit forms of~$\mathcal{L}(k^2)$ and~$\mathcal{R}(p)$}
\label{sec:A}

We first define the propagators
\begin{align}
   \Hat{h}_{\mu\nu}(p)
   &=e^{-2p^2}
   \left[
   \left(\delta_{\mu\nu}-\frac{1}{p^2}p_\mu p_\nu\right)
   \frac{1}{e^{-2p^2}+p^2\mathcal{K}_\tau(p^2)}
   +\frac{1}{p^2}p_\mu p_\nu
   \frac{\xi_\tau}{\xi_\tau e^{-2p^2}+p^2}
   \right],
\notag\\
   \Hat{h}_F(p)
   &=e^{-2p^2}\frac{i}{e^{-2p^2}+i(\Slash{p}+im_\tau)}.
\label{eq:(A1)}
\end{align}
We also define the vertex functions
$\Hat{V}_{\mu_1\dotsb\mu_n}(-p-k_1-\dotsb-k_n,p;k_1,\dotsc,k_n)$ as the coefficients
in the expansion of the fermion part of the 1PI action in powers of the gauge
potential:
\begin{align}
   &\left.\mathit{\Gamma}_\tau\right|_{O(\Bar{\Psi}\Psi)}
\notag\\
   &=-\int_p\,
   \Bar{\Psi}_{-1}(-p)(\Slash{p}+im_\tau)\Psi_{-1}(p)
   +e_\tau\int_{p,k}\,
   \Bar{\Psi}_{-1}(-p-k)e^{k^2}\Slash{\mathcal{A}}(k)\Psi_{-1}(p)
\notag\\
   &=-\int_p\,\Bar{\Psi}(-p)\Hat{V}(p)\Psi(p)
\notag\\
   &\qquad{}
   +\sum_{n=1}^\infty e_\tau^n\int_{p,k_1,\dotsc,k_n}
   \Bar{\Psi}(-p-k_1-\dotsb-k_n)
\notag\\
   &\qquad\qquad{}
   \times
   \Hat{V}_{\mu_1\dotsb\mu_n}(-p-k_1-\dotsb-k_n,p;k_1,\dotsc,k_n)
   \mathcal{A}_{\mu_1}(k_1)\dotsb\mathcal{A}_{\mu_n}(k_n)
   \Psi(p),
\label{eq:(A2)}
\end{align}
where
\begin{equation}
   \Hat{V}(p):=e^{p^2}(\Slash{p}+im_\tau)e^{p^2},
\label{eq:(A3)}
\end{equation}
which is related to the fermion propagator by~$\Hat{h}_F(p)^{-1}=\Hat{V}(p)-i$,
and the explicit form of~$\Hat{V}_{\mu_1\dotsb\mu_n}$ is studied
in~Appendix~\ref{sec:B}.

We then define the following combinations:
\begin{equation}
   \Hat{\mathcal{A}}_{\mu\nu}
   :=2\Hat{V}_{\mu\nu}+\Hat{V}_{\mu}\Hat{h}_F\Hat{V}_{\nu}+\Hat{V}_{\nu}\Hat{h}_F\Hat{V}_{\mu},
\label{eq:(A4)}
\end{equation}
\begin{align}
   \Hat{\mathcal{B}}_{\mu\nu\rho}
   &:=6\Hat{V}_{\mu\nu\rho}
\notag\\
   &\qquad{}
   +2\Hat{V}_{\mu\nu}\Hat{h}_F\Hat{V}_\rho
   +2\Hat{V}_{\nu\rho}\Hat{h}_F\Hat{V}_\mu
   +2\Hat{V}_{\rho\mu}\Hat{h}_F\Hat{V}_\nu
   +2\Hat{V}_\mu \Hat{h}_F\Hat{V}_{\nu\rho}
   +2\Hat{V}_\nu \Hat{h}_F\Hat{V}_{\rho\mu}
   +2\Hat{V}_\rho \Hat{h}_F\Hat{V}_{\mu\nu}
\notag\\
   &\qquad{}
   +\Hat{V}_\mu \Hat{h}_F\Hat{V}_\nu \Hat{h}_F\Hat{V}_\rho+\Hat{V}_\nu \Hat{h}_F\Hat{V}_\rho \Hat{h}_F\Hat{V}_\mu+\Hat{V}_\rho \Hat{h}_F\Hat{V}_\mu \Hat{h}_F\Hat{V}_\nu
\notag\\
   &\qquad{}
   +\Hat{V}_\mu \Hat{h}_F\Hat{V}_\rho \Hat{h}_F\Hat{V}_\nu+\Hat{V}_\rho \Hat{h}_F\Hat{V}_\nu \Hat{h}_F\Hat{V}_\mu+\Hat{V}_\nu \Hat{h}_F\Hat{V}_\mu \Hat{h}_F\Hat{V}_\rho,
\label{eq:(A5)}
\end{align}
and
\begin{align}
   \Hat{\mathcal{C}}_{\mu\nu\rho\sigma}
   &:=24\Hat{V}_{\mu\nu\rho\sigma}
\notag\\
   &\qquad{}
   +6\Hat{V}_{\mu\nu\rho}\Hat{h}_F\Hat{V}_{\sigma}+6\Hat{V}_{\nu\rho\sigma}\Hat{h}_F\Hat{V}_{\mu}
   +6\Hat{V}_{\rho\sigma\mu}\Hat{h}_F\Hat{V}_{\nu}+6\Hat{V}_{\sigma\mu\nu}\Hat{h}_F\Hat{V}_{\rho}
\notag\\
   &\qquad{}
   +6\Hat{V}_{\mu}\Hat{h}_F\Hat{V}_{\nu\rho\sigma}+6\Hat{V}_{\nu}\Hat{h}_F\Hat{V}_{\rho\sigma\mu}
   +6\Hat{V}_{\rho}\Hat{h}_F\Hat{V}_{\sigma\mu\nu}+6\Hat{V}_{\sigma}\Hat{h}_F\Hat{V}_{\mu\nu\rho}
\notag\\
   &\qquad{}
   +4\Hat{V}_{\mu\nu}\Hat{h}_F\Hat{V}_{\rho\sigma}+4\Hat{V}_{\nu\rho}\Hat{h}_F\Hat{V}_{\sigma\mu}+4\Hat{V}_{\rho\sigma}\Hat{h}_F\Hat{V}_{\mu\nu}
\notag\\
   &\qquad{}
   +4\Hat{V}_{\sigma\mu}\Hat{h}_F\Hat{V}_{\nu\rho}+4\Hat{V}_{\mu\rho}\Hat{h}_F\Hat{V}_{\nu\sigma}+4\Hat{V}_{\nu\sigma}\Hat{h}_F\Hat{V}_{\mu\rho}
\notag\\
   &\qquad{}
   +2\Hat{V}_{\mu\nu}\Hat{h}_F\Hat{V}_{\rho}\Hat{h}_F\Hat{V}_{\sigma}+2\Hat{V}_{\nu\rho}\Hat{h}_F\Hat{V}_{\sigma}\Hat{h}_F\Hat{V}_{\mu}
   +2\Hat{V}_{\rho\sigma}\Hat{h}_F\Hat{V}_{\mu}\Hat{h}_F\Hat{V}_{\nu}+2\Hat{V}_{\sigma\mu}\Hat{h}_F\Hat{V}_{\nu}\Hat{h}_F\Hat{V}_{\rho}
\notag\\
   &\qquad{}
   +2\Hat{V}_{\mu\nu}\Hat{h}_F\Hat{V}_{\sigma}\Hat{h}_F\Hat{V}_{\rho}+2\Hat{V}_{\nu\rho}\Hat{h}_F\Hat{V}_{\mu}\Hat{h}_F\Hat{V}_{\sigma}
   +2\Hat{V}_{\rho\sigma}\Hat{h}_F\Hat{V}_{\nu}\Hat{h}_F\Hat{V}_{\mu}+2\Hat{V}_{\sigma\mu}\Hat{h}_F\Hat{V}_{\rho}\Hat{h}_F\Hat{V}_{\nu}
\notag\\
   &\qquad{}
   +2\Hat{V}_{\mu\rho}\Hat{h}_F\Hat{V}_{\nu}\Hat{h}_F\Hat{V}_{\sigma}+2\Hat{V}_{\mu\rho}\Hat{h}_F\Hat{V}_{\sigma}\Hat{h}_F\Hat{V}_{\nu}
   +2\Hat{V}_{\nu\sigma}\Hat{h}_F\Hat{V}_{\mu}\Hat{h}_F\Hat{V}_{\rho}+2\Hat{V}_{\nu\sigma}\Hat{h}_F\Hat{V}_{\rho}\Hat{h}_F\Hat{V}_{\mu}
\notag\\
   &\qquad{}
   +2\Hat{V}_{\mu}\Hat{h}_F\Hat{V}_{\nu\rho}\Hat{h}_F\Hat{V}_{\sigma}+2\Hat{V}_{\nu}\Hat{h}_F\Hat{V}_{\rho\sigma}\Hat{h}_F\Hat{V}_{\mu}
   +2\Hat{V}_{\rho}\Hat{h}_F\Hat{V}_{\sigma\mu}\Hat{h}_F\Hat{V}_{\nu}+2\Hat{V}_{\sigma}\Hat{h}_F\Hat{V}_{\mu\nu}\Hat{h}_F\Hat{V}_{\rho}
\notag\\
   &\qquad{}
   +2\Hat{V}_{\mu}\Hat{h}_F\Hat{V}_{\rho\sigma}\Hat{h}_F\Hat{V}_{\nu}+2\Hat{V}_{\mu}\Hat{h}_F\Hat{V}_{\sigma\nu}\Hat{h}_F\Hat{V}_{\rho}
   +2\Hat{V}_{\nu}\Hat{h}_F\Hat{V}_{\sigma\mu}\Hat{h}_F\Hat{V}_{\rho}+2\Hat{V}_{\nu}\Hat{h}_F\Hat{V}_{\mu\rho}\Hat{h}_F\Hat{V}_{\sigma}
\notag\\
   &\qquad{}
   +2\Hat{V}_{\rho}\Hat{h}_F\Hat{V}_{\mu\nu}\Hat{h}_F\Hat{V}_{\sigma}+2\Hat{V}_{\rho}\Hat{h}_F\Hat{V}_{\nu\sigma}\Hat{h}_F\Hat{V}_{\mu}
   +2\Hat{V}_{\sigma}\Hat{h}_F\Hat{V}_{\nu\rho}\Hat{h}_F\Hat{V}_{\mu}+2\Hat{V}_{\sigma}\Hat{h}_F\Hat{V}_{\rho\mu}\Hat{h}_F\Hat{V}_{\nu}
\notag\\
   &\qquad{}
   +2\Hat{V}_{\mu}\Hat{h}_F\Hat{V}_{\nu}\Hat{h}_F\Hat{V}_{\rho\sigma}+2\Hat{V}_{\nu}\Hat{h}_F\Hat{V}_{\rho}\Hat{h}_F\Hat{V}_{\sigma\mu}
   +2\Hat{V}_{\rho}\Hat{h}_F\Hat{V}_{\sigma}\Hat{h}_F\Hat{V}_{\mu\nu}+2\Hat{V}_{\sigma}\Hat{h}_F\Hat{V}_{\mu}\Hat{h}_F\Hat{V}_{\nu\rho}
\notag\\
   &\qquad{}
   +2\Hat{V}_{\mu}\Hat{h}_F\Hat{V}_{\rho}\Hat{h}_F\Hat{V}_{\sigma\nu}+2\Hat{V}_{\mu}\Hat{h}_F\Hat{V}_{\sigma}\Hat{h}_F\Hat{V}_{\nu\rho}
   +2\Hat{V}_{\nu}\Hat{h}_F\Hat{V}_{\sigma}\Hat{h}_F\Hat{V}_{\mu\rho}+2\Hat{V}_{\nu}\Hat{h}_F\Hat{V}_{\mu}\Hat{h}_F\Hat{V}_{\rho\sigma}
\notag\\
   &\qquad{}
   +2\Hat{V}_{\rho}\Hat{h}_F\Hat{V}_{\mu}\Hat{h}_F\Hat{V}_{\nu\sigma}+2\Hat{V}_{\rho}\Hat{h}_F\Hat{V}_{\nu}\Hat{h}_F\Hat{V}_{\sigma\mu}
   +2\Hat{V}_{\sigma}\Hat{h}_F\Hat{V}_{\nu}\Hat{h}_F\Hat{V}_{\rho\mu}+2\Hat{V}_{\sigma}\Hat{h}_F\Hat{V}_{\rho}\Hat{h}_F\Hat{V}_{\mu\nu}
\notag\\
   &\qquad{}
   +\Hat{V}_{\mu}\Hat{h}_F\Hat{V}_{\nu}\Hat{h}_F\Hat{V}_{\rho}\Hat{h}_F\Hat{V}_{\sigma}+\Hat{V}_{\mu}\Hat{h}_F\Hat{V}_{\rho}\Hat{h}_F\Hat{V}_{\sigma}\Hat{h}_F\Hat{V}_{\nu}
   +\Hat{V}_{\mu}\Hat{h}_F\Hat{V}_{\sigma}\Hat{h}_F\Hat{V}_{\nu}\Hat{h}_F\Hat{V}_{\rho}
\notag\\
   &\qquad{}
   +\Hat{V}_{\mu}\Hat{h}_F\Hat{V}_{\nu}\Hat{h}_F\Hat{V}_{\sigma}\Hat{h}_F\Hat{V}_{\rho}+\Hat{V}_{\mu}\Hat{h}_F\Hat{V}_{\rho}\Hat{h}_F\Hat{V}_{\nu}\Hat{h}_F\Hat{V}_{\sigma}
   +\Hat{V}_{\mu}\Hat{h}_F\Hat{V}_{\sigma}\Hat{h}_F\Hat{V}_{\rho}\Hat{h}_F\Hat{V}_{\nu}
\notag\\
   &\qquad{}
   +\Hat{V}_{\nu}\Hat{h}_F\Hat{V}_{\rho}\Hat{h}_F\Hat{V}_{\sigma}\Hat{h}_F\Hat{V}_{\mu}+\Hat{V}_{\nu}\Hat{h}_F\Hat{V}_{\sigma}\Hat{h}_F\Hat{V}_{\mu}\Hat{h}_F\Hat{V}_{\rho}
   +\Hat{V}_{\nu}\Hat{h}_F\Hat{V}_{\mu}\Hat{h}_F\Hat{V}_{\rho}\Hat{h}_F\Hat{V}_{\sigma}
\notag\\
   &\qquad{}
   +\Hat{V}_{\nu}\Hat{h}_F\Hat{V}_{\rho}\Hat{h}_F\Hat{V}_{\mu}\Hat{h}_F\Hat{V}_{\sigma}+\Hat{V}_{\nu}\Hat{h}_F\Hat{V}_{\sigma}\Hat{h}_F\Hat{V}_{\rho}\Hat{h}_F\Hat{V}_{\mu}
   +\Hat{V}_{\nu}\Hat{h}_F\Hat{V}_{\mu}\Hat{h}_F\Hat{V}_{\sigma}\Hat{h}_F\Hat{V}_{\rho}
\notag\\
   &\qquad{}
   +\Hat{V}_{\rho}\Hat{h}_F\Hat{V}_{\sigma}\Hat{h}_F\Hat{V}_{\mu}\Hat{h}_F\Hat{V}_{\nu}+\Hat{V}_{\rho}\Hat{h}_F\Hat{V}_{\mu}\Hat{h}_F\Hat{V}_{\nu}\Hat{h}_F\Hat{V}_{\sigma}
   +\Hat{V}_{\rho}\Hat{h}_F\Hat{V}_{\nu}\Hat{h}_F\Hat{V}_{\sigma}\Hat{h}_F\Hat{V}_{\mu}
\notag\\
   &\qquad{}
   +\Hat{V}_{\rho}\Hat{h}_F\Hat{V}_{\sigma}\Hat{h}_F\Hat{V}_{\nu}\Hat{h}_F\Hat{V}_{\mu}+\Hat{V}_{\rho}\Hat{h}_F\Hat{V}_{\mu}\Hat{h}_F\Hat{V}_{\sigma}\Hat{h}_F\Hat{V}_{\nu}
   +\Hat{V}_{\rho}\Hat{h}_F\Hat{V}_{\nu}\Hat{h}_F\Hat{V}_{\mu}\Hat{h}_F\Hat{V}_{\sigma}
\notag\\
   &\qquad{}
   +\Hat{V}_{\sigma}\Hat{h}_F\Hat{V}_{\mu}\Hat{h}_F\Hat{V}_{\nu}\Hat{h}_F\Hat{V}_{\rho}+\Hat{V}_{\sigma}\Hat{h}_F\Hat{V}_{\nu}\Hat{h}_F\Hat{V}_{\rho}\Hat{h}_F\Hat{V}_{\mu}
   +\Hat{V}_{\sigma}\Hat{h}_F\Hat{V}_{\rho}\Hat{h}_F\Hat{V}_{\mu}\Hat{h}_F\Hat{V}_{\nu}
\notag\\
   &\qquad{}
   +\Hat{V}_{\sigma}\Hat{h}_F\Hat{V}_{\mu}\Hat{h}_F\Hat{V}_{\rho}\Hat{h}_F\Hat{V}_{\nu}+\Hat{V}_{\sigma}\Hat{h}_F\Hat{V}_{\nu}\Hat{h}_F\Hat{V}_{\mu}\Hat{h}_F\Hat{V}_{\rho}
   +\Hat{V}_{\sigma}\Hat{h}_F\Hat{V}_{\rho}\Hat{h}_F\Hat{V}_{\nu}\Hat{h}_F\Hat{V}_{\mu}.
\label{eq:(A6)}
\end{align}

Using these combinations, we define
\begin{align}
   &e^{2k^2}\mathcal{R}_{\mu\nu}^{\text{loop}}(k)
\notag\\
   &:=ie_\tau^2\int_\ell\,
   (-4\ell^2+2\gamma_\psi-1)
   \tr\left[\Hat{h}_F(\ell)\Hat{\mathcal{A}}_{\mu\nu}\Hat{h}_F(\ell)\right]
\notag\\
   &\qquad{}
   +4ie_\tau^2\int_\ell\,
   \tr\left[\Hat{h}_F(\ell-k)\Hat{V}_\mu\Hat{h}_F(\ell)\right]
   (2\ell-k)_\nu
\notag\\
   &\qquad{}
   +4ie_\tau^2\int_\ell\,(2\ell+k)_\mu
   \tr\left[\Hat{h}_F(\ell+k)\Hat{V}_\nu\Hat{h}_F(\ell)\right]
\notag\\
   &\qquad{}
   -8ie_\tau^2\delta_{\mu\nu}\int_\ell\,
   \tr\left[\Hat{h}_F(\ell)\right]
\notag\\
   &\qquad{}
   +4ie_\tau^4\int_{\ell,\ell'}\,
   (2\ell_\alpha+\ell_\alpha')\Hat{h}_{\alpha\beta}(\ell')
   \tr\left[\Hat{h}_F(\ell+\ell')\Hat{\mathcal{B}}_{\beta\mu\nu}\Hat{h}_F(\ell)
   \right]
\notag\\
   &\qquad{}
   -8ie_\tau^4\int_{\ell,\ell'}\,\Hat{h}_{\beta\nu}(\ell')
   \tr\left[\Hat{h}_F(\ell)\Hat{\mathcal{A}}_{\beta\mu}\Hat{h}_F(\ell+\ell'+k)
   \right]
\notag\\
   &\qquad{}
   -8ie_\tau^4\int_{\ell,\ell'}\,\Hat{h}_{\beta\mu}(\ell')
   \tr\left[\Hat{h}_F(\ell)\Hat{\mathcal{A}}_{\beta\nu}\Hat{h}_F(\ell+\ell'-k)
   \right]
\notag\\
   &\qquad{}
   -4ie_\tau^4\int_\ell\,\Hat{h}_{\rho\rho}(\ell)
   \int_{\ell'}\,
   \tr\left[\Hat{h}_F(\ell')\Hat{\mathcal{A}}_{\mu\nu}\Hat{h}_F(\ell')\right]
\notag\\
   &\qquad{}
   -4ie_\tau^6\int_{\ell,\ell',\ell''}\,
   \Hat{h}_{\alpha\beta}(\ell')
   \Hat{h}_{\alpha\gamma}(\ell'')
   \tr\left[
   \Hat{h}_F(\ell+\ell'+\ell'')\Hat{\mathcal{C}}_{\beta\gamma\mu\nu}\Hat{h}_F(\ell)
   \right].
\label{eq:(A7)}
\end{align}
As shown in~I, this $\mathcal{R}_{\mu\nu}^{\text{loop}}(k)$ is transverse,
$k_\mu\mathcal{R}_{\mu\nu}^{\text{loop}}(k)=0$, as implied by the gauge WT identity.
Therefore, we can set $\mathcal{R}_{\mu\nu}^{\text{loop}}(k)=(k^2\delta_{\mu\nu}-k_\mu k_\nu)\mathcal{L}(k^2)$ and
\begin{equation}
   \mathcal{L}(k^2)
   =\frac{1}{(D-1)k^2}\mathcal{R}_{\mu\mu}^{\text{loop}}(k).
\label{eq:(A8)}
\end{equation}
This is the function~$\mathcal{L}(k^2)$ in~Eq.~\eqref{eq:(2.13)}.

The function $\mathcal{R}(p)$ in~Eq.~\eqref{eq:(2.14)} is, on the other hand,
given by
\begin{align}
   &e^{2p^2}\mathcal{R}(p)
\notag\\
   &:=-e_\tau^2\int_\ell\,
   (2\ell^2+1-\epsilon-\gamma_A)\Hat{h}_{\mu\rho}(\ell)\Hat{h}_{\sigma\mu}(\ell)
   \Hat{\mathcal{A}}_{\rho\sigma}
\notag\\
   &\qquad{}
   +ie_\tau^2\int_\ell\,
   \left[4(p+\ell)^2+1-2\gamma_\psi\right]
   \Hat{h}_{\mu\nu}(\ell)\Hat{V}_\mu \Hat{h}_F(p+\ell)^2\Hat{V}_\nu
\notag\\
   &\qquad{}
   -4ie_\tau^2\int_\ell\,
   \Hat{h}_{\mu\nu}(\ell)p_\mu \Hat{h}_F(p+\ell)\Hat{V}_\nu
\notag\\
   &\qquad{}
   -4e_\tau^2\int_\ell\,
   \Hat{h}_{\mu\nu}(\ell)\Hat{V}_\mu \Hat{h}_F(p+\ell)\Hat{V}(p)(p+\ell)_\nu
\notag\\
   &\qquad{}
   -4ie_\tau^2\int_\ell\,
   \Hat{h}_{\mu\nu}(\ell)\Hat{V}_\mu \Hat{h}_F(p+\ell)p_\nu
\notag\\
   &\qquad{}
   -4e_\tau^2\int_\ell\,
   \Hat{h}_{\mu\nu}(\ell)(p+\ell)_\mu\Hat{V}(p)\Hat{h}_F(p+\ell)\Hat{V}_\nu
\notag\\
   &\qquad{}
   +4e_\tau^2\int_\ell\,
   \Hat{h}_{\mu\mu}(\ell)\Hat{V}(p)
\notag\\
   &\qquad{}
   -4ie_\tau^4
   \int_{\ell,\ell'}\,(2p+\ell+2\ell')_\mu
   \Hat{h}_{\mu\nu}(-\ell)\Hat{h}_{\rho\sigma}(+\ell')
   \Hat{\mathcal{A}}_{\nu\rho}
   \Hat{h}_F(p+\ell+\ell')\Hat{h}_F(p+\ell')\Hat{V}_\sigma
\notag\\
   &\qquad{}
   -4ie_\tau^4
   \int_{\ell,\ell'}\,(2p+\ell+2\ell')_\mu
   \Hat{h}_{\mu\nu}(+\ell)\Hat{h}_{\rho\sigma}(-\ell')
   \Hat{V}_\sigma \Hat{h}_F(p+\ell')\Hat{h}_F(p+\ell+\ell')
   \Hat{\mathcal{A}}_{\nu\rho}
\notag\\
   &\qquad{}
   +8ie_\tau^4
   \int_{\ell,\ell'}\,\ell_\mu
   \Hat{h}_{\mu\nu}(+\ell')\Hat{h}_{\rho\sigma}(+\ell')
   \tr[\Hat{h}_F(\ell+\ell')\Hat{V}_\sigma \Hat{h}_F(\ell)]
   \Hat{\mathcal{A}}_{\nu\rho}
\notag\\
   &\qquad{}
   +8ie_\tau^4
   \int_{\ell,\ell'}\,
   \Hat{h}_{\mu\nu}(-\ell)\Hat{h}_{\rho\mu}(-\ell')
   \Hat{V}_\nu \Hat{h}_F(p+\ell)\Hat{h}_F(p+\ell')\Hat{V}_\rho
\notag\\
   &\qquad{}
   +4ie_\tau^4
   \int_{\ell,\ell'}\,
   \Hat{h}_{\mu\mu}(+\ell)
   \Hat{h}_{\rho\sigma}(+\ell')
   \Hat{V}_\rho \Hat{h}_F(p+\ell')^2\Hat{V}_\sigma
\notag\\
   &\qquad{}
   -4ie_\tau^4
   \int_{\ell,\ell'}\,
   \tr[\Hat{h}_F(+\ell)]
   \Hat{h}_{\mu\nu}(-\ell')\Hat{h}_{\rho\mu}(-\ell')\Hat{\mathcal{A}}_{\nu\rho}
\notag\\
   &\qquad{}
   +2ie_\tau^4
   \int_{\ell,\ell'}\,
   \Hat{h}_{\mu\nu}(+\ell)\Hat{h}_{\rho\mu}(-\ell')
   \Hat{h}_F(p+\ell+\ell')\Hat{\mathcal{A}}_{\nu\rho}
\notag\\
   &\qquad{}
   +2ie_\tau^4
   \int_{\ell,\ell'}\,
   \Hat{h}_{\mu\nu}(-\ell)\Hat{h}_{\rho\mu}(+\ell')
   \Hat{\mathcal{A}}_{\nu\rho}
   \Hat{h}_F(p+\ell+\ell')
\notag\\
   &\qquad{}
   +2e_\tau^4
   \int_{\ell,\ell'}\,
   \Hat{h}_{\mu\nu}(-\ell)\Hat{h}_{\rho\mu}(+\ell')
   \Hat{\mathcal{A}}_{\nu\rho}
   \Hat{h}_F(p+\ell+\ell')\Hat{V}(p)
\notag\\
   &\qquad{}
   +2e_\tau^4
   \int_{\ell,\ell'}\,
   \Hat{h}_{\mu\nu}(+\ell)\Hat{h}_{\rho\mu}(-\ell')\Hat{V}(p)
   \Hat{h}_F(p+\ell+\ell')\Hat{\mathcal{A}}_{\nu\rho}
\notag\\
   &\qquad{}
   +4ie_\tau^6\int_{\ell,\ell',\ell''}\,
   \Hat{h}_{\mu\nu}(+\ell)\Hat{h}_{\mu\rho}(+\ell')\Hat{h}_{\sigma\tau}(-\ell'')
   \Hat{V}_\tau \Hat{h}_F(p+\ell'')\Hat{h}_F(p+\ell+\ell'+\ell'')
   \Hat{\mathcal{B}}_{\rho\sigma\nu}
\notag\\
   &\qquad{}
   +4ie_\tau^6\int_{\ell,\ell',\ell''}\,
   \Hat{h}_{\mu\nu}(-\ell)\Hat{h}_{\mu\rho}(-\ell')\Hat{h}_{\sigma\tau}(-\ell'')
   \Hat{\mathcal{B}}_{\nu\rho\tau}
   \Hat{h}_F(p+\ell+\ell'+\ell'')
   \Hat{h}_F(p+\ell'')\Hat{V}_\sigma
\notag\\
   &\qquad{}
   -4ie_\tau^6\int_{\ell,\ell',\ell''}\,
   \Hat{h}_{\mu\nu}(+\ell)\Hat{h}_{\mu\rho}(+\ell')\Hat{h}_{\sigma\tau}(+\ell+\ell')
   \tr[\Hat{h}_F(\ell+\ell'+\ell'')\Hat{V}_\tau \Hat{h}_F(\ell'')]
   \Hat{\mathcal{B}}_{\rho\sigma\nu}
\notag\\
   &\qquad{}
   +8ie_\tau^6\int_{\ell,\ell',\ell''}\,
   \Hat{h}_{\mu\nu}(-\ell)\Hat{h}_{\mu\rho}(+\ell')\Hat{h}_{\sigma\tau}(-\ell'')
   \Hat{\mathcal{A}}_{\tau\nu}\Hat{h}_F(p+\ell+\ell'')\Hat{h}_F(p+\ell'+\ell'')
   \Hat{\mathcal{A}}_{\rho\sigma}
\notag\\
   &\qquad{}
   -8ie_\tau^6\int_{\ell,\ell',\ell''}\,
   \Hat{h}_{\mu\nu}(+\ell)\Hat{h}_{\mu\rho}(+\ell')\Hat{h}_{\sigma\tau}(+\ell')
   \tr[\Hat{h}_F(\ell+\ell'+\ell'')\Hat{\mathcal{A}}_{\tau\nu}\Hat{h}_F(\ell'')]
   \Hat{\mathcal{A}}_{\rho\sigma}.
\label{eq:(A9)}
\end{align}

In the above expressions, we have adopted the following abbreviation rules for
the assignment of momentum variables. First, when propagators are attached to a
vertex function, the momentum variables of the vertex function are identified
with the momenta of the attached propagators. For example,
\begin{equation}
   \Hat{h}_{\rho\mu}(k)\Hat{h}_F(-p)\Hat{V}_{\mu\nu\dotsb}\Hat{h}_F(q)
   =\Hat{h}_{\rho\mu}(k)\Hat{h}_F(-p)
   \Hat{V}_{\mu\nu\dotsb}(p,q;k,\dotsc)\Hat{h}_F(q).
\label{eq:(A10)}
\end{equation}
Similarly, when the index of the vertex function is the first index of the
photon propagator, the sign of the momentum is reversed:
\begin{equation}
   \Hat{h}_{\mu\rho}(k)\Hat{h}_F(-p)\Hat{V}_{\mu\nu\dotsb}\Hat{h}_F(q)
   =\Hat{h}_{\mu\rho}(k)\Hat{h}_F(-p)
   \Hat{V}_{\mu\nu\dotsb}(p,q;-k,\dotsc)\Hat{h}_F(q).
\label{eq:(A11)}
\end{equation}
 Note that, since $\Hat{h}_{\mu\nu}(k)$ is symmetric in~$\mu$ and~$\nu$ and even
in~$k$, the order of the indices and the sign of the argument
in~$\Hat{h}_{\mu\nu}$ serve only to specify the momentum flowing into the vertex.
In Eq.~\eqref{eq:(A7)}, the momenta associated with the indices~$\mu$ and~$\nu$
are understood to be~$-k$ and~$k$, respectively; in Eq.~\eqref{eq:(A9)}, the
momenta of the external fermion lines are $-p$ (left) and~$p$ (right).
Finally, arguments of functions not fixed by the above rules are determined by
momentum conservation at the vertex. For instance, in the first term
of~Eq.~\eqref{eq:(A7)},
\begin{align}
   &\Hat{h}_F(\ell)\Hat{\mathcal{A}}_{\mu\nu}\Hat{h}_F(\ell)
\notag\\
   &=\Hat{h}_F(\ell)\Hat{\mathcal{A}}_{\mu\nu}(-\ell,\ell;-k,k)\Hat{h}_F(\ell)
\notag\\
   &=\Hat{h}_F(\ell)
   \Bigl[
   2\Hat{V}_{\mu\nu}(-\ell,\ell;-k,k)
   +\Hat{V}_\mu(-\ell,\ell+k;-k)\Hat{h}_F(\ell+k)\Hat{V}_\nu(-\ell-k,\ell;k)
\notag\\
   &\qquad\qquad{}
   +\Hat{V}_\nu(-\ell,\ell-k;k)\Hat{h}_F(\ell-k)\Hat{V}_\mu(-\ell+k,\ell;-k)
   \Bigr]  
   \Hat{h}_F(\ell).
\label{eq:(A12)}
\end{align}
By these rules, the assignment of the momentum variables is uniquely determined.

\section{Vertex functions associated with the $-1$ variables}
\label{sec:B}
In this appendix, we derive the vertex functions~$\Hat{V}_{\mu_1\dotsb\mu_n}$
defined by~Eq.~\eqref{eq:(A2)} to all orders in~$e_\tau$.

For~$\alpha_0=1$, the solution of the first equation in~Eq.~\eqref{eq:(2.9)} is
$\mathcal{A}_{t\mu}(k)=e^{-tk^2}\mathcal{A}_\mu(k)$, and in particular
$\mathcal{A}_{-1\mu}(k)=e^{k^2}\mathcal{A}_\mu(k)$. Setting
$\Psi_t(p)=e^{-tp^2}\varphi_t(p)$, the flow equation for the fermion
field in~Eq.~\eqref{eq:(2.9)} becomes
\begin{align}
   \partial_t\varphi_t(p)
   &=2e\int_k\,e^{t[p^2-k^2-(p-k)^2]}\mathcal{A}_\mu(k)(p-k)_\mu\varphi_t(p-k)
\notag\\
   &\qquad{}
   -e^2\int_{k,l}\,e^{t[p^2-k^2-l^2-(p-k-l)^2]}
   \mathcal{A}_\mu(k)\mathcal{A}_\mu(l)\varphi_t(p-k-l),
\label{eq:(B1)}
\end{align}
with the initial condition~$\varphi_0=\Psi$; then
$\Psi_{-1}(p)=e^{p^2}\varphi_{-1}(p)$. Solving Eq.~\eqref{eq:(B1)} backward from
$t=0$ to~$t=-1$ by
iterating~$\varphi_{-1}=\varphi_0-\int_{-1}^0dt\,\partial_t\varphi_t$, each term
is given by an integral over a simplex of the exponential of a linear function
of the flow times. This can be expressed by the divided difference of the
exponential function through the Hermite--Genocchi formula,
\begin{equation}
   \exp[q_0,q_1,\dotsc,q_m]
   =\int_{0\leq s_j,\sum_js_j=1}d^ms\,\exp\left(\sum_{j=0}^ms_jq_j\right),
\label{eq:(B2)}
\end{equation}
where the divided difference is defined recursively by $f[q_0]=f(q_0)$
and~$f[q_0,\dotsc,q_m]=(f[q_1,\dotsc,q_m]-f[q_0,\dotsc,q_{m-1}])/(q_m-q_0)$; it
is symmetric in the nodes~$q_j$.

The result for the $-1$ variable is
\begin{equation}
   \Psi_{-1}(p)
   =\sum_{n\geq0}e^n\int_{k_1,\dotsc,k_n}\,
   \Hat{w}_n(p;k_1,\dotsc,k_n)_{\mu_1\dotsb\mu_n}
   \mathcal{A}_{\mu_1}(k_1)\dotsb\mathcal{A}_{\mu_n}(k_n)
   \Psi(p-k_1-\dotsb-k_n),
\label{eq:(B3)}
\end{equation}
where
\begin{equation}
   \Hat{w}_n(p;k_1,\dotsc,k_n)
   =\sum_{\substack{c_j\in\{1,2\}\\n=c_1+\dotsb+c_m}}
   \exp[q_0,q_1,\dotsc,q_m]
   \prod_{j=1}^m
   \begin{cases}
   -2(P_j)_\mu&(c_j=1),\\
   \delta_{\mu\nu}&(c_j=2).
   \end{cases}
\label{eq:(B4)}
\end{equation}
Here, the sum is over the ordered partitions of~$n$ into~$1$ and~$2$, and the
$j$th block carries the next $c_j$ photons. We have defined
\begin{align}
   &P_0=p,\qquad
   P_j=P_{j-1}-(\text{sum of the photon momenta in the $j$th block}),
\notag\\
   &q_j=P_j^2+\sum_{\text{photons in blocks $1,\dotsc,j$}}k^2,
\label{eq:(B5)}
\end{align}
that is, $q_j$ is the sum of the squared momentum of the fermion remaining
after the $j$th block and the squared momenta of the photons attached so far.
For example,
\begin{equation}
   \Hat{w}_0(p)=e^{p^2},\qquad
   \Hat{w}_1(p;k)_\mu
   =-2(p-k)_\mu\frac{e^{(p-k)^2+k^2}-e^{p^2}}{(p-k)^2+k^2-p^2}.
\label{eq:(B6)}
\end{equation}
The expression of~$\Bar{\Psi}_{-1}$ is obtained by~$e\to-e$.

Substituting these into the first line of~Eq.~\eqref{eq:(A2)}, we have
\begin{align}
   &\Tilde{\Hat{V}}(k_1,\dotsc,k_n)
\notag\\
   &=\sum_{a=0}^n(-1)^a\Hat{w}_a(-P;k_1,\dotsc,k_a)
   (-1)(\Slash{P}+im_\tau)
   \Hat{w}_{n-a}(P;k_{a+1},\dotsc,k_n)
\notag\\
   &\qquad{}
   +\sum_{a=0}^{n-1}(-1)^a\Hat{w}_a(-P-k_{a+1};k_1,\dotsc,k_a)
   \gamma_{\mu_{a+1}}e^{k_{a+1}^2}\,
   \Hat{w}_{n-a-1}(P;k_{a+2},\dotsc,k_n),
\label{eq:(B7)}
\end{align}
and
\begin{equation}
   \Hat{V}_{\mu_1\dotsb\mu_n}
   =\frac{1}{n!}\sum_{\sigma\in S_n}
   \Tilde{\Hat{V}}(k_{\sigma(1)},\dotsc,k_{\sigma(n)}),
\label{eq:(B8)}
\end{equation}
where, in each term, $P$ is $p$ plus the sum of the photon momenta in the
factor~$\Hat{w}$ on the right, and the indices are permuted together with the
momenta. The factor~$e^{k_{a+1}^2}$ in the second line of~Eq.~\eqref{eq:(B7)}
arises from $\mathcal{A}_{-1\mu}=e^{k^2}\mathcal{A}_\mu$ in the interaction
term. The $n=0$ term,
$-\Hat{w}_0(-P)(\Slash{P}+im_\tau)\Hat{w}_0(P)=-\Hat{V}(P)$, reproduces
Eq.~\eqref{eq:(A3)}. The Dirac structure of the vertex functions is always a
linear combination of~$\Slash{P}+im_\tau$ and~$\gamma_\mu$. The cases $n=1$
and~$2$ reproduce Eqs.~(A2) and~(A4)--(A6) of~I, where the latter are related
to~$\Hat{V}_{\mu_1\dotsb\mu_n}$ by the factors $e^{(p+\sum k)^2}$, $e^{p^2}$,
and~$e^{k_i^2}$ associated with the variables used in~I [Eq.~(3.25) of~I].



%



\let\doi\relax










\end{document}